\documentclass[reprint,pra,amsmath,amssymb,aps,showpacs,preprintnumbers]{revtex4-2}
\usepackage{graphicx} % Include figure files
\usepackage{datetime}
\usepackage{bm} % bold math

\begin{document}
\title{Spontaneous decay of a quantum emitter near a plasmonic nanostructure}
%\title{Spontaneous emission  mediated by cooperative energy transfer to plasmonic antenna}
\author{Tigran V. Shahbazyan}
\affiliation{
Department of Physics, Jackson State University, Jackson, MS
39217 USA}

%\date{\today,\,\,\currenttime}

\begin{abstract} 
We develop a theory for spontaneous decay of a quantum emitter (QE) situated near metal-dielectric structure supporting localized surface plasmons. If plasmon resonance is tuned close to the  QE emission frequency, the   emission  is enhanced due to energy transfer from the QE to a localized plasmon mode followed by photon emission by plasmonic antenna. The emission rate is determined by intimate interplay between the plasmon coupling to  radiation field and the Ohmic losses in metal.  Here we develop plasmon Green's function approach that includes plasmon's interaction with radiation to obtain explicit expressions for radiative decay rate and  optical polarizability of a localized plasmon mode in arbitrary plasmonic nanostructure. Within this approach, we provide   consistent definition of  plasmon mode volume  by relating it to plasmon mode density, which characterizes the plasmon field confinement, and recover the standard cavity form of the  Purcell factor, but now for plasmonic systems. We show that, for QE placed at a  "hot spot" near a sharp tip of a small metal nanostructure, the plasmon mode volume scales with the metal volume while being very sensitive to the proximity to the tip. Finally, we derive the enhancement factor for radiated power spectrum for any nanoplasmonic system and relate it to the Purcell factor for spontaneous decay rate. We illustrate our results by numerical  example of a QE situated near  gold nanorod tip.
\end{abstract}
%\pacs{ 78.67.-n, 78.67.Bf}
\maketitle

%%%%%%%%%%%%%%%%%%%%%%%%%%%%%%%%%%%%%%%%%%%%

%
%%%%%%%%%%%%%%%%%%%%%%%%%%%%%%%%%%%%%%%%%%
\section{Introduction}
\label{sec:intro}
Rapid advances in nanoplasmonics during past decade opened up avenues for extremely high energy concentration and transfer on   length scale well below  the diffraction limit \cite{atwater-jap05,ozbay-science06,stockman-review}. Optical interactions between  dye molecules or semiconductor quantum dots, hereafter reffered to as quantum emitters (QEs),  and localized plasmons in metal-dielectric  structures  underpin major phenomena in plasmon-enhanced spectroscopy, such as surface-enhanced Raman scattering (SERS) \cite{sers}, plasmon-enhanced fluorescence and luminescence \cite{feldmann-prl02,artemyev-nl02,novotny-prl06,sandoghdar-prl06,halas-nl07,halas-acsnano09,ming-nl09}, strong QE-plasmon coupling \cite{bellessa-prl04,sugawara-prl06,wurtz-nl07,fofang-nl08,hakala-prl09,gomez-nl10,manjavacas-nl11,berrier-acsnano11,salomon-prl12,guebrou-prl12,antosiewicz-acsphotonics14,luca-apl14}, and plasmonic laser \cite{bergman-prl03,stockman-natphot08,noginov-nature09}.  On the theory side, however, despite significant progress in various  aspects of plasmonics, a consistent description of spontaneous decay of a QE  near a plasmonic nanostructure  characterized by dispersive and lossy dielectric function is still  a subject of active debate \cite{carminati-oc06,greffet-prl10,lalanne-prl13,hughes-njp14,pelton-np15,belov-sr15,derex-jo16,greffet-acsph17,koenderink-acsphot17,lalanne-prb18}.

Spontaneous decay of a QE near a photonic or plasmonic resonator can be strongly modified due to additional energy transfer (ET) channel provided by the QE coupling to cavity or plasmonic modes \cite{novotny-book}. If the mode frequency  $\omega_{m}$ is tuned close to the QE emission frequency, the QE decay rate can be greatly enhanced relative to the  free-space decay rate $\gamma_{0}^{r}$. The modified rate   is usually presented as   $\gamma=\gamma_{0}^{r}+\gamma_{et}=\gamma_{0}^{r}\left (1+F_{p}\right ) $, where   $\gamma_{et}$ is  the ET rate between QE and resonant mode  whereas $F_{p}$ is the Purcell factor characterizing the decay rate enhancement \cite{purcell-pr46}. For QE coupled to cavity mode, the Purcell factor has the form
\begin{equation}
\label{purcell}
F_{p}=\dfrac{\gamma_{et}}{\gamma_{0}^{r}}=\frac{6\pi Q_{m}}{k^{3}{\cal V}_{m}},
\end{equation}
where $Q_{m}$ is the mode quality factor, ${\cal V}_{m}$ is the mode volume and $k=\omega/c$ is the light wave vector ($\omega$ and $c$ are frequency and speed of light). For photonic cavities, the mode volume  at some point $\bm{r}$ is defined as ${\cal V}_{cav}=\int dV \varepsilon(\bm{r}) |\bm{E}_{m}(\bm{r})|^{2} /[\varepsilon(\bm{r}) |\bm{E}_{m}(\bm{r})|^{2}]$, where $\bm{E}_{m}(\bm{r})$ is the mode  electric field and $\varepsilon(\bm{r})$ is (lossless) dielectric function, and is usually interpreted as the volume that would confine the mode at given field intensity.

Spontaneous decay of a QE coupled to \textit{plasmonic} resonator has  been addressed within several approaches \cite{lalanne-prl13,hughes-njp14,pelton-np15,belov-sr15,derex-jo16,greffet-acsph17,koenderink-acsphot17,lalanne-prb18}   aiming to obtain  the corresponding Purcell factor in the form (\ref{purcell}). While  the plasmon quality factor is well defined as $Q_{m}=\omega_{m}/\gamma_{m}$, where  $\gamma_{m}$ is the plasmon decay rate, there has been active debate as to how unambiguously define the plasmon mode volume for  QE   located outside a metal nanostructure characterized by complex dispersive dielectric function \cite{maier-oe06,koenderink-ol10,stefano-jpcm10,andreani-prb12,hughes-ol12,lalanne-pra14,hughes-acsphot14,hughes-pra15,muljarov-prb16,bergman-17,lalanne-prb18,lalanne-lpr18}. For open systems, straightforward analogies with photonic cavities  do not apply  and more rigorous, albeit less intuitive, numerical methods based on modal expansion of Maxwell equations' solutions are often employed \cite{hughes-acsphot14,lalanne-lpr18}.

Here we develop another approach more suitable for nanoplasmonic systems which extends the quasistatic approximation, valid for system scale  below the diffraction limit, to incorporate  the plasmon coupling to the radiation field in a consistent way. Specifically, if the system size $L$ is much smaller than the photon wavelength $\lambda $  then, on the far-field scale $r \gg \lambda$,   interaction of  localized plasmon  mode  with radiation field is analogous to that of a point-like emitter with dipole moment $\bm{\mathcal{P}}_{m}=\int dV \bm{P}_{m}$, where $\bm{P}_{m}(\bm{r})$ is the  electric polarization vector of the plasmon mode.  On the other hand, on the near field scale $L\ll \lambda$,   QE decay involves ET to  plasmon  at a rate $\gamma_{et}$   determined by the plasmon local density of states (LDOS) \cite{shahbazyan-prl16}. Subsequently, some part of transferred energy is  radiated away by the plasmonic antenna while the rest is dissipated  in metal due to the Ohmic losses. An accurate treatment of spontaneous decay requires matching the balance between transferred and dissipated energy  in the near field  to the radiated energy in the far field. As we show in this paper, this is accomplished by including the plasmon coupling to radiation field into the plasmon Green's function, which defines the LDOS, in a  way that ensures  energy flux conservation across the scales.

In the preceding paper \cite{shahbazyan-prl16}, we derived the plasmon Green's function for arbitrary metal-dielectric system with Ohmic losses included, but \textit{without} coupling to the radiation field, in order to describe plasmonic enhancement of Forster ET between  donors and  acceptors. In this paper, we extend our approach to include the plasmon coupling to  radiation field, and derive explicit expression for  the plasmon radiative decay rate $\gamma_{m}^{r}$. By incorporating $\gamma_{m}^{r}$ into the plasmon Green's function, we obtain optical polarizability of plasmonic system describing its response to an external field in the way that satisfies energy flux conservation.  We then turn to spontaneous decay of a QE  coupled to plasmonic resonator and derive the Purcell factor for decay rate in the form (\ref{purcell}), where the mode volume is identified as the inverse of plasmon mode density that characterizes  plasmon field confinement at the QE position. We show that  near sharp tip  of small metal nanostructure, where  the plasmon field is strongly confined (hot spot),  the mode volume scales with the metal volume but, at the same time, is very sensitive to the QE distance to metallic tip. Finally, we derive enhancement factor  for radiated power spectrum, which describes, e.g., plasmonic enhancement of fluorescence near metal nanostructures \cite{feldmann-prl02,artemyev-nl02,novotny-prl06,sandoghdar-prl06,halas-nl07,halas-acsnano09,ming-nl09},  and establish general relation between the enhancement and Purcell factors.

The paper is organized as follows. In Sec.~\ref{sec2} we revisit our derivation of the plasmon Green's function \cite{shahbazyan-prl16} by using different method that makes its generalization more convenient. In Sec.~\ref{sec3}, we extend this approach by including the plasmon coupling to  radiation field into Green's function, and derive explicit expressions for radiative decay rate and optical polarization of any nanoplasmonic system. In Sec.~\ref{sec4}, we derive the plasmon LDOS,   plasmon mode density, and    plasmon mode volume, as well as evaluate the plasmon mode volume near sharp tip of  metal nanostructure. In Sec.~\ref{sec5}, we derive the Purcell factor for spontaneous decay of a QE coupled to plasmonic resonator  and   obtain explicit expression for the power spectrum enhancement factor. In Sec.~\ref{sec6}, we illustrate our results numerically for a QE situated neat the tip of Au nanorod. A summary of our results is provided in Sec.~\ref{sec7}, and some details of our calculations are outlined in the appendix.

\section{Spontaneous decay and plasmon Green's function }
\label{sec2}

Consider an excited QE  with  dipole matrix element and orientation $\mu$ and $\bm{n}$, respectively, located at some position $\bm{r}$ near   metal-dielectric structure described by the complex  dielectric function  $\varepsilon (\omega,\bm{r})=\varepsilon' (\omega,\bm{r})+i\varepsilon'' (\omega,\bm{r})$ and  surrounded by a homogeneous medium with dielectric constant $\varepsilon_{s}$. We set $\varepsilon_{s}=1$ for now, but will restore it when discussing numerical examples. The full decay rate  of a QE in electromagnetic environment has the form \cite{novotny-book}
\begin{equation}
\label{rate-em}
\gamma =  \frac{8\pi \omega^{2}\mu^{2}}{c^{2}\hbar}\,\text{Im} \left [ \bm{n}\!\cdot\!\bar{\bm{G}}(\omega;\bm{r},\bm{r})\!\cdot\! \bm{n}\right ],
\end{equation}
where $\bar{\bm{G}}(\omega;\bm{r},\bm{r}')$ is the dyadic Green function for Maxwell's equation satisfying $\bm{\nabla}\times \bm{\nabla}\times \bar{\bm{G}}-(\omega^{2}/c^{2})\varepsilon\, \bar{\bm{G}} = \bm{I}$. For a  QE in free space, the decay rate is determined by the imaginary part of the free-space Green function $\bar{\bm{G}}_{0}(\omega;\bm{r},\bm{r}')$  at the QE position,  $\text{Im}[\bar{\bm{G}}_{0}(\omega;\bm{r},\bm{r})]=(\omega/6\pi c)\bm{I}$, yielding 
%$\gamma_{0}^{r}=4\mu^{2}\omega^{3}/3\hbar c^{3}$ \cite{novotny-book}.
%
\begin{equation}
\label{rate-free}
\gamma_{0}^{r}=\dfrac{4\mu^{2}\omega^{3}}{3\hbar c^{3}}.
\end{equation}
For systems with characteristic  size below the diffraction limit, it is convenient to use the  rescaled Green's function,
\begin{equation}
\label{dyadic-nearfield}
\bar{\bm{D}}(\omega;\bm{r},\bm{r}')=\frac{4\pi\omega^{2}}{c^{2}}\,\bar{\bm{G}}(\omega;\bm{r},\bm{r}'),
\end{equation}
which, in the \textit{near-field} limit,  represents the sum of direct and plasmon terms,  $\bar{\bm{D}}=\bar{\bm{D}}_{0}+\bar{\bm{D}}_{\rm pl}$ \cite{shahbazyan-prl16}. The full decay rate (\ref{rate-em}) takes the form $\gamma =\gamma_{0}^{r}+\gamma_{et}$, where
\begin{equation}
\label{rate-et}
\gamma_{et}=\frac{2 \mu^{2}}{\hbar}\,\text{Im} \left [ \bm{n}\!\cdot\!\bar{\bm{D}}_{\rm pl}(\omega;\bm{r},\bm{r})\!\cdot\! \bm{n}\right ]
\end{equation}
is QE-plasmon ET rate.

\subsection{Plasmon Green's function: Lossless case}

For a metal-dielectric system with characteristic size smaller than the radiation wavelength, the fields and frequencies of plasmon modes  are determined by  the quasistatic Gauss law \cite{stockman-review}
\begin{equation}
\label{gauss}
\bm{\nabla}\!\cdot\!\left [\varepsilon' (\omega_{m},\bm{r})\bm{\nabla}\Phi_{m}(\bm{r})\right ]=0,
\end{equation}
where  the potentials $\Phi_{m}(\bm{r})$, which  define  the mode electric fields as $\bm{E}_{m}(\bm{r})=-\bm{\nabla}\Phi_{m}(\bm{r})$, satisfy the standard boundary conditions across metal-dielectric interfaces. The mode  fields, which we chose to be real, are orthogonal, $\int\! dV \bm{E}_{m}(\bm{r})\!\cdot\!\bm{E}_{n}(\bm{r})=\delta_{mn}\int\! dV \bm{E}_{m}^{2}(\bm{r})$, and  regular inside  the structure while falling off  rapidly  outside it.

In our preceeding paper \cite{shahbazyan-prl16}, the plasmon Green function in the presence of Ohmic losses  was derived  by expressing it through complex eigenvalues of the  operator $\bm{\nabla}\!\cdot\!\left [\varepsilon (\omega,\bm{r})\bm{\nabla}\right ]$. In this section, we give a more transparent derivation  without resorting to an eigenvalue problem, which permits its generalization to include, in the next section, the plasmon coupling to a radiation field.

The Green's function $S(\omega;\bm{r},\bm{r}')$ for quasistatic potentials satisfies the equation
\begin{equation}
\label{gauss-green}
\bm{\nabla}\!\cdot\!\left [\varepsilon (\omega,\bm{r})\bm{\nabla}S(\omega;\bm{r},\bm{r}')\right ]=4\pi \delta(\bm{r}-\bm{r}'),
\end{equation}
for arbitrary frequency $\omega$. In free space ($\varepsilon=1$), the quasistatic Green's function is independent of frequency and has the form $S_{0}(\bm{r}-\bm{r}')=-1/|\bm{r}-\bm{r}'|$; the corresponding dyadic Green's function for fields, given by  $\bm{\nabla}\bm{\nabla}'S_{0}(\bm{r}-\bm{r}')$, coincides with  (the real part of) free-space electromagnetic Green's function (\ref{dyadic-nearfield}) in the near-field limit. After splitting  $S$ into free-space and plasmon parts, $S=S_{0}+S_{\rm pl}$, we obtain an equation for $S_{\rm pl}$:
\begin{align}
\label{gauss-green-plas}
\bm{\nabla}\!\cdot\!\bigl[\varepsilon (\omega,\bm{r})\bm{\nabla}
&S_{\rm pl}(\omega;\bm{r},\bm{r}')\bigr]
\nonumber\\
&
=
-\bm{\nabla}\!\cdot\!\bigl [[\varepsilon (\omega,\bm{r})-1]\bm{\nabla}S_{0}(\omega;\bm{r},\bm{r}')\bigr ].
\end{align}
Assume, for a moment, that  the dielectric function $\varepsilon (\omega,\bm{r})$ is lossless ($\varepsilon''=0$). For real $\varepsilon$, the Green's function can be expanded in terms of eigenmodes of Eq.~(\ref{gauss}) as 
\begin{equation}
\label{green-exp}
S_{\rm pl}(\omega;\bm{r},\bm{r}')=\sum_{m}S_{m}(\omega)\Phi_{m}(\bm{r})\Phi_{m}(\bm{r}'),
\end{equation}
where coefficients $S_{m}(\omega)$ are found as follows. Applying to Eq.~(\ref{gauss-green-plas})  the integral operator $\int\! dV'\Phi_{m}(\bm{r}')\Delta'$, and using    the relation
\begin{equation}
\int\! dV'\Phi_{m}(\bm{r}')\Delta'S_{\rm pl}(\omega;\bm{r},\bm{r}')= - S_{m} \Phi_{m}(\bm{r}) \!\int\! dV \bm{E}_{m}^{2}(\bm{r}) 
\end{equation}
for the left-hand side, and the relation
\begin{equation}
\int\! dV'\Phi_{m}(\bm{r}')\Delta'S_{0}(\omega;\bm{r},\bm{r}')= 4\pi\Phi_{m}(\bm{r})
\end{equation}
for the right-hand side, we obtain
\begin{equation}
\label{gauss-green-plas2}
S_{m} \bm{\nabla}\!\cdot\!\bigl[\varepsilon (\omega,\bm{r})\bm{\nabla}
\Phi_{m}(\bm{r})\bigr]=  
4\pi \dfrac{\bm{\nabla}\!\cdot\!\bigl [[\varepsilon (\omega,\bm{r})-1]\bm{\nabla}\Phi_{m}(\bm{r})\bigr ]}{\int\! dV \bm{E}_{m}^{2}(\bm{r})}.
\end{equation}
Finally, multiplying Eq.~(\ref{gauss-green-plas2}) by $\Phi_{m}(\bm{r})$ and integrating  the result over $\bm{r}$, we obtain
\begin{equation}
\label{gauss-green-plas3}
S_{m}(\omega)= \dfrac{4\pi}{\int\! dV \bm{E}_{m}^{2}(\bm{r})} -  \dfrac{4\pi}{\int\! dV \varepsilon (\omega,\bm{r})\bm{E}_{m}^{2}(\bm{r})}.
\end{equation}
For \textit{real} $\varepsilon(\omega,\bm{r})$, the Green function (\ref{green-exp}) with coefficients (\ref{gauss-green-plas3}) is \textit{exact} for any metal-dielectric structure with eigenmodes defined by Eq.~(\ref{gauss}). The first  term in Eq.~(\ref{gauss-green-plas3}) ensures that $S_{m}=0$ in the limit $\omega\rightarrow\infty$ (or, in free space with $\varepsilon=1$), while the second term develops a pole, due to the Gauss law (\ref{gauss}), as $|\omega|$ approaches $\omega_{m}$.

\subsection{Plasmon Green's function: Including the losses}

For a complex dielectric function, the plasmon poles in the Green's function move into the lower half of the complex-frequency plane. We assume that the mode quality factors $Q_{m}$ are sufficiently large  and so,  in the first order in $1/Q_{m}$, the eigenmodes $\Phi_{m}$ in the Green's function expansion (\ref{green-exp}) are unchanged while the coefficients $S_{m}$  in Eq.~(\ref{gauss-green-plas3}) are now complex. The higher-order corrections come from the "dissipation coupling" between the modes $\int\! dV \varepsilon'' (\omega,\bm{r})\bm{E}_{m}(\bm{r})\cdot\bm{E}_{n}(\bm{r})$. Upon expanding  the dielectric function near $\omega_{m}$,
\begin{equation}
\varepsilon(\omega,\bm{r})\approx \varepsilon' (\omega_{m},\bm{r})+ \dfrac{ \partial \varepsilon' (\omega_{m},\bm{r})}{\partial \omega_{m}^{2}}\left (\omega^{2}-\omega_{m}^{2}\right ) + i\varepsilon'' (\omega,\bm{r}),
\end{equation}
the coefficients (\ref{gauss-green-plas3})  take the form
\begin{equation}
\label{mode-coeff}
S_{m}(\omega) =\frac{\omega_{m}^{2}}{2U_{m}}\dfrac{1}{\omega_{m}^{2}-\omega^{2} -i\omega\gamma_{m}^{nr}(\omega)},
\end{equation}
where 
\begin{align}
\label{energy-mode}
U_{m}
&= \frac{\omega_{m}}{16\pi} 
\!\int \!  dV     
\dfrac{\partial\varepsilon'(\omega_{m},\bm{r})}{\partial \omega_{m}}
 \bm{E}_{m}^{2}(\bm{r}) 
 \nonumber\\
&= \frac{1}{16\pi} 
\!\int \!  dV     
\dfrac{\partial[\omega_{m}\varepsilon'(\omega_{m},\bm{r})]}{\partial \omega_{m}}
 \bm{E}_{m}^{2}(\bm{r}) 
%\partial [\omega_{m}\varepsilon'(\omega_{m},\bm{r})]/\partial \omega_{m},
\end{align}
is the plasmon mode energy \cite{landau},  and the rate
\begin{equation}
\label{mode-decay-bos-nr0}
\gamma_{m}^{nr}(\omega)=\dfrac{2\omega_{m}\!\int \!  dV     
\varepsilon''(\omega,\bm{r})\bm{E}_{m}^{2}(\bm{r}) }{\omega \!\int \!  dV     
[\partial\varepsilon'(\omega_{m},\bm{r})/\partial \omega_{m}]
 \bm{E}_{m}^{2}(\bm{r}) }
\end{equation}
describes \textit{nonradiative} plasmon decay at frequency $\omega$. Introducing the power dissipated by the plasmon mode due to nonradiative (Ohmic) losses as \cite{landau}
\begin{equation}
\label{mode-power-nr}
W_{m}^{nr}(\omega)=\frac{\omega}{8\pi}\!\int \! dV  \varepsilon''(\omega,\bm{r})  \bm{E}_{m}^{2}(\bm{r}),
\end{equation}
the frequency-dependent nonradiative plasmon decay rate (\ref{mode-decay-bos-nr0}) can be written in the form
\begin{equation}
\label{mode-decay-bos-nr}
\gamma_{m}^{nr}(\omega)= \dfrac{\omega_{m}^{2}}{\omega^{2}}\dfrac{W_{m}^{nr}(\omega)}{U_{m}},
\end{equation}
which is convenient for extension in the next section. 

The  quasistatic dyadic Green's function for the electric fields is given by $\bar{\bm{D}}_{\rm pl}(\omega;\bm{r},\bm{r}')=\bm{\nabla}\bm{\nabla}' S_{\rm pl}(\omega;\bm{r},\bm{r}')$, where $S_{\rm pl}(\omega;\bm{r},\bm{r}')$ is given by Eq.~(\ref{green-exp}) with coefficients $S_{m}(\omega)$ given by Eq.~(\ref{mode-coeff}), and has the form
\begin{equation}
\label{dyadic-mode-bos}
\bar{\bm{D}}_{\rm pl}(\omega;\bm{r},\bm{r}') = \sum_{m}\frac{\omega_{m}^{2}}{2 U_{m}}\frac{\bm{E}_{m}(\bm{r}) \bm{E}_{m}  (\bm{r}')}{\omega_{m}^{2}-\omega^{2} -i\omega\gamma_{m}^{nr}(\omega)}.
\end{equation}
Note that the coefficients (\ref{mode-coeff})  are obtained by calculating the residues at the plasmon poles of function $S_{m}(\omega)$, given by Eq.~(\ref{gauss-green-plas3}), and the Green's function (\ref{dyadic-mode-bos}) is obtained by summing up the contributions from  all poles. Since the plasmon Green's function is analytic in the complex-frequency plane except isolated poles in the lower half plane [for local dielectric function $\varepsilon(\omega,\bm{r})$],  the expression (\ref{dyadic-mode-bos})  is valid for \textit{all} frequencies. The functional form of the decay rate (\ref{mode-decay-bos-nr0}) along with the modes' orthogonality ensures that $\bar{\bm{D}}_{\rm pl}(\omega;\bm{r},\bm{r}')$ obeys the optical theorem \cite{optical}
\begin{align}
\label{optical}
\int \! dV  \varepsilon''(\omega,\bm{r}) \bar{\bm{D}}_{\rm pl}^{*}(\omega;\bm{r},\bm{r}')
&\bar{\bm{D}}_{\rm pl}(\omega;\bm{r},\bm{r}'')
\nonumber\\
&=4\pi \text{Im}\bar{\bm{D}}_{\rm pl}(\omega;\bm{r}',\bm{r}''),
\end{align}
which, in the absence of radiation, implies that  the system's energy intake (right-hand side) is dissipated via Ohmic losses  (left-hand side) \cite{poddubny-prl16}.

In the following, we assume that the QE's interaction with the plasmonic system is dominated by a single mode and, accordingly, keep only the resonant term in Eq.~(\ref{dyadic-mode-bos}),
\begin{align}
\label{dyadic-mode}
\bar{\bm{D}}_{m}(\omega;\bm{r},\bm{r}') 
=\frac{\omega_{m}^{2}}{2 U_{m}}\frac{\bm{E}_{m}(\bm{r}) \bm{E}_{m}  (\bm{r}')}{\omega_{m}^{2}-\omega^{2} -i\omega\gamma_{m}(\omega)},
\end{align}
where $\gamma_{m}(\omega)=\gamma_{m}^{nr}(\omega)$ for the quasistatic case. For a well-defined plasmon mode, i.e., if the quality factor is sufficiently large ($\omega_{m}/\gamma_{m}\gg 1$), the contribution from negative frequencies is small  and the plasmon Green's function near the resonance takes the form \cite{shahbazyan-prl16} 
\begin{align}
\label{dyadic-mode-res}
\bar{\bm{D}}_{m}(\omega;\bm{r},\bm{r}') 
=\frac{\omega_{m}}{4 U_{m}}\frac{\bm{E}_{m}(\bm{r}) \bm{E}_{m}  (\bm{r}')}{\omega_{m}-\omega-i\gamma_{m}/2 },
\end{align}
where  $\gamma_{m}=W_{m}/U_{m}$ is the plasmon decay rate at the plasmon frequency [with $W_{m}\equiv W_{m}^{nr}(\omega_{m})$ in the quasistatic case]. Note that single-mode Green functions (\ref{dyadic-mode}) and (\ref{dyadic-mode-res}) also satisfy the optical theorem (\ref{optical}) (the latter with $\omega=\omega_{m}$). Finally, since only metallic regions with the dispersive dielectric function $\varepsilon(\omega)=\varepsilon'(\omega)+i\varepsilon''(\omega)$ contribute to $U_{m}$ and $W_{m}^{nr}$,   the  standard plasmon decay rate due to nonradiative losses in metal is recovered,
\begin{equation}
\label{mode-rate-nr}
\gamma_{m}^{nr}
%=\frac{W_{m}^{nr}}{U_{m}}
=\frac{2\varepsilon''(\omega_{m})}{\partial \varepsilon'(\omega_{m})/\partial \omega_{m}}.
\end{equation}
In the next section, we generalize our approach to include the plasmon  interaction with the radiation field.

\section{Interaction of plasmon mode with radiation field}

\label{sec3}

In this section, we demonstrate that  the quasistatic  Green's function (\ref{dyadic-mode}) can be extended to incorporate  the plasmon coupling to the radiation field by including plasmon's \textit{radiative} decay rate into a full decay rate as follows [compare with Eq.~(\ref{mode-decay-bos-nr})]:
\begin{equation}
\label{mode-decay-bos}
\gamma_{m}(\omega)=\dfrac{\omega_{m}^{2}}{\omega^{2}}\frac{W_{m}(\omega)}{U_{m}},
\end{equation}
where $W_{m}(\omega)=W_{m}^{nr}(\omega)+W_{m}^{r}(\omega)$ is the \textit{full} dissipated power, which now includes  the radiated power  $W_{m}^{r}(\omega)$ that determines the plasmon's radiative decay rate as 
\begin{equation}
\label{mode-decay-bos-rad}
\gamma_{m}^{r}(\omega)=\dfrac{\omega_{m}^{2}}{\omega^{2}}\frac{W_{m}^{r}(\omega)}{U_{m}}.
\end{equation}
Below, we derive explicit expressions for the radiated power $W_{m}^{r}(\omega)$  as well as for the optical polarizability of a plasmon mode characterizing a plasmonic system's response to an external field.

\subsection{Radiative decay of plasmon mode}

We start by noting that emission of light from a plasmonic system with characteristic size much smaller than the radiation wavelength can be treated similarly to  a point dipole. The frequency-dependent polarization vector of  plasmon mode (\ref{gauss}) is $\bm{P}_{m}(\omega,\bm{r})= \chi'(\omega,\bm{r}) \bm{E}_{m}(\bm{r})$, where $ \chi(\omega,\bm{r}) = \left [\varepsilon(\omega,\bm{r})-1\right ]/4\pi$ is the plasmonic system's susceptibility that vanishes outside the system (we assume, for simplicity, that the dielectric constant of the surrounding medium is unity). Note that,  in the plasmon  spectral domain $\varepsilon''(\omega)/\varepsilon'(\omega)\ll 1$, the radiation and scattering by a plasmonic dipole are determined, within our approximation, by the real part of the susceptibility $\chi'= (\varepsilon'-1)/4\pi$ whereas its imaginary part $\chi''=\varepsilon''/4\pi$ determines the Ohmic losses (\ref{mode-power-nr}). The electric field generated by the plasmonic system's oscillating polarization vector is given by
\begin{equation}
\label{mode-field-far}
\bm{\mathcal{E}}_{m}(\omega,\bm{r})=\int \! dV' \, \bar{\bm{D}}_{0}(\omega;\bm{r},\bm{r}')\!\cdot\! \bm{P}_{m}(\omega,\bm{r}'),
\end{equation}
where $\bar{\bm{D}}_{0}(\omega;\bm{r},\bm{r}')=(4\pi\omega^{2}/c^{2})\bar{\bm{G}}_{0}(\omega;\bm{r},\bm{r}')$ is the  free-space dyadic Green's function. The power dissipated by the plasmon mode via radiation is given by  \cite{novotny-book}
\begin{align}
W_{m}^{r}(\omega)
&=\dfrac{\omega}{2}\,\text{Im}\!\int\! dV \bm{\mathcal{E}}_{m}(\omega,\bm{r})\!\cdot\! \bm{P}_{m}(\omega,\bm{r})
\\
=&\dfrac{\omega}{2}\,\text{Im}\!\int\! dV \!\int\! dV'  \bm{P}_{m}(\omega,\bm{r}) \!\cdot\! \bar{\bm{D}}_{0}(\omega;\bm{r},\bm{r}')\!\cdot\! \bm{P}_{m}(\omega,\bm{r}'),
\nonumber
\end{align}
where integration takes place over the plasmonic system volume. Replacing the free-space Green's function by its near-field limit, $\text{Im}\bar{\bm{D}}_{0}(\omega;\bm{r},\bm{r}')=(2\omega^{3}/3c^{3})\bm{I}$, we obtain 
\begin{equation}
\label{mode-power-rad}
W_{m}^{r}(\omega)
%= \frac{\omega^{4}}{3c^{3}} \left |\int \! dV  \bm{P}_{m}(\bm{r})\right |^{2}
= \frac{\omega^{4}}{3c^{3}} \,\bm{\mathcal{P}}_{m}^{2}(\omega),
\end{equation}
where 
\begin{equation}
\label{mode-dipole}
\bm{\mathcal{P}}_{m}(\omega)
=\int \! dV  \bm{P}_{m}(\omega,\bm{r})
= \frac{1}{4\pi}\!\int \! dV \left [\varepsilon'(\omega,\bm{r})-1\right ]\bm{E}_{m}(\bm{r})
\end{equation}
is the plasmon's dipole moment.  The same result is obtained by integrating the  Poynting's  vector $S=(c/8\pi)\left |\bm{\mathcal{E}}_{m}(\omega,\bm{r})\right |^{2}$ over remote surface enclosing the   system. Note that plasmon's radiated power (\ref{mode-power-rad}) coincides with that of a point dipole $\bm{\mathcal{P}}_{m}(\omega)$, and that, for small systems,  radiation of higher-order multipoles is suppressed \cite{novotny-book}. By including the  radiated power (\ref{mode-power-rad}) into the full dissipated power,  the radiative decay channel is incorporated, through the decay rate (\ref{mode-decay-bos}), within  the plasmon Green's function (\ref{dyadic-mode}), in a way that ensures energy flux conservation (see below).

Near the plasmon resonance, the plasmon decay rate in the Green's function (\ref{dyadic-mode-res}) takes the form $\gamma_{m}=\gamma_{m}^{nr}+\gamma_{m}^{r}$, where the plasmon radiation rate is obtained by normalizing the radiated power with  the mode energy, 
\begin{equation}
\label{mode-rate-rad}
\gamma_{m}^{r}=\dfrac{W_{m}^{r}}{U_{m}}
=
\frac{\omega_{m}^{4}}{3c^{3}} \frac{\bm{\mathcal{P}}_{m}^{2}}{U_{m}}, 
\end{equation}
which, upon using Eqs.~(\ref{energy-mode}) and (\ref{mode-dipole}),  takes the form
\begin{equation}
\label{mode-rate-rad2}
\gamma_{m}^{r}
=\frac{\omega_{m}^{4}}{3\pi c^{3}} 
\,
\frac{\left [ \int \! dV  (\varepsilon'-1)\bm{E}_{m}(\bm{r})\right ]^{2}}{ \int \! dV(\partial \omega_{m} \varepsilon'/\partial \omega_{m})\bm{E}_{m}^{2}(\bm{r})},
\end{equation}
where we denoted $W_{m}^{r}\equiv W_{m}^{r}(\omega_{m})$, $\bm{\mathcal{P}}_{m}\equiv \bm{\mathcal{P}}_{m}(\omega_{m})$   and, under the integral, $\varepsilon\equiv \varepsilon(\omega_{m},\bm{r})$. Correspondingly, the plasmon radiation efficiency $\eta_{m}$ has the form
\begin{equation}
\label{mode-efficiency}
\eta_{m}=\frac{\gamma_{m}^{r}}{\gamma_{m}}=\dfrac{\zeta_{m}}{1+\zeta_{m}},
\end{equation}
where the parameter
\begin{equation}
\label{mode-rate-relative}
\zeta_{m}=\frac{\gamma_{m}^{r}}{\gamma_{m}^{nr}}
=\frac{\omega_{m}^{3}}{6\pi c^{3}} 
\,
\frac{\left [ \int \! dV  (\varepsilon'-1)\bm{E}_{m}(\bm{r})\right ]^{2}}{ \int \! dV \varepsilon'' \bm{E}_{m}^{2}(\bm{r})},
\end{equation}
characterizes the plasmon's radiative decay rate vs its nonradiative decay rate.  Note that, for small nanoplasmonic systems,  $\gamma_{m}^{nr}$ should also include the  Landau damping rate  \cite{shahbazyan-prb16}.  

As an example, for a dipole surface plasmon in a spherical nanoparticle of radius $a$,  a straightforward  calculation  recovers  the radiative decay rate as (see appendix) 
\begin{equation}
\label{mode-rate-rad-sp}
\gamma_{sp}^{r}= \frac{4\omega_{sp}^{3}a^{3}}{c^{3}\partial \varepsilon'( \omega_{sp})/\partial \omega_{sp}},
\end{equation}
and, correspondingly,   $\zeta_{sp}=2\omega_{sp}^{3}a^{3}/c^{3}\varepsilon''( \omega_{sp})$, where the plasmon frequency $\omega_{sp}$ is given by $\varepsilon'(\omega_{sp})=-2$.

Finally note that, in contrast to a field-independent nonradiative decay rate (\ref{mode-rate-nr}), the radiative decay rate (\ref{mode-rate-rad2}) \textit{does} depend on the plasmon field  distribution in the system, albeit not on  its overall magnitude. Such "nonanalytic" field dependence of $\gamma_{m}^{r}$, which is present in the Landau damping rate as well \cite{shahbazyan-prb16}, reflects the fact that, in contrast to a point dipole, the  local fields vary appreciably on the plasmonic system's scale.

\subsection{Optical polarizability of a plasmonic system in the external field and energy flux conservation}

Here we show that the plasmon Green's function that incorporates the Ohmic and radiation losses ensures the standard relation between a plasmon's absorption, scattering and extinction cross sections, $\sigma_{abs}+\sigma_{sc}=\sigma_{ext}$, and derive the optical polarizability of the plasmon mode which describes the plasmonic system's resonant response to an external field.  For a frequency close to the plasmon resonance, we use the single-mode plasmon Green function (\ref{dyadic-mode}) and, accordingly, omit nonresonant contributions.

\subsubsection{Extinction cross section and energy flux conservation}

Consider the response of a plasmonic system to an incident   monochromatic field $\bm{\mathcal{E}}_{i}e^{-i\omega t}$ that is uniform on the system scale. The electric field scattered by the plasmonic system has the form
\begin{equation}
\label{mode-field-sc}
\bm{\mathcal{E}}_{\rm sc}(\omega,\bm{r})=\int \! dV'  \chi'(\omega,\bm{r}')\bar{\bm{D}}(\omega;\bm{r},\bm{r}')\!\cdot\! \bm{\mathcal{E}}_{i},
\end{equation}
where  $\bar{\bm{D}}(\omega;\bm{r},\bm{r}')$ is the dyadic  Green function (\ref{dyadic-nearfield}). The power absorbed by the plasmonic structure is
\begin{equation}
\label{mode-power-abs}
P_{\rm abs}(\omega)=\frac{\omega}{8\pi}\!\int \! dV  \varepsilon''(\omega,\bm{r})  |\bm{\mathcal{E}}_{\rm sc}(\omega,\bm{r})|^{2},
\end{equation}
where we disregarded nonresonant direct field absorption. Inside the plasmonic system, for each mode, we replace $\bar{\bm{D}}(\omega;\bm{r},\bm{r}')$ in Eq.~(\ref{mode-field-sc}) with the plasmon Green's function $\bar{\bm{D}}_{m}(\omega;\bm{r},\bm{r}')$, given by Eq.~(\ref{dyadic-mode}), and  obtain 
\begin{equation}
P_{\rm abs}(\omega)=W_{m}^{nr}(\omega)|S_{m}(\omega)|^{2} [\bm{\mathcal{P}}_{m}(\omega)\!\cdot\!\bm{\mathcal{E}}_{i}]^{2}, 
\end{equation}
where the functions $S_{m}(\omega)$, $W_{m}^{nr}(\omega)$,  and $\bm{\mathcal{P}}_{m}(\omega)$ are given by Eqs.~(\ref{mode-coeff}), (\ref{mode-power-nr}) and (\ref{mode-dipole}), respectively. Normalizing $P_{\rm abs}(\omega)$ by the  incident energy flux $S_{i}=(c/8\pi)\bm{\mathcal{E}}_{i}^{2}$, we obtain the mode \textit{absorption} cross section
\begin{equation}
\label{cross-abs}
\sigma_{\rm abs}^{(m)}(\omega)=\dfrac{4\pi\omega}{c}\dfrac{\omega_{m}^{2}}{2U_{m}}\dfrac{\omega\gamma_{m}^{nr}(\omega)\,[\bm{e}\!\cdot\!\bm{\mathcal{P}}_{m}(\omega)]^{2}}{(\omega_{m}^{2}-\omega^{2})^{2}+\omega^{2}\gamma_{m}^{2}(\omega)},
\end{equation}
where the plasmon decay rates $\gamma_{m}^{nr}(\omega)$  and  $\gamma_{m}(\omega)$ are given by Eqs.~(\ref{mode-decay-bos-nr}) and  (\ref{mode-decay-bos}), respectively, and the unit vector $\bm{e}$ is the incident field polarization.

To obtain the scattering cross section, we extract the far-field contribution from Eq.~(\ref{mode-field-sc}) with the help of the Dyson equation for the dyadic  Green's function,
\begin{align}
\label{dyson}
\bar{\bm{D}}(\omega;\bm{r},\bm{r}')
&=\bar{\bm{D}}_{0}(\omega;\bm{r},\bm{r}')\\
&+\int \! dV_{1} \chi'(\omega,\bm{r}_{1})\bar{\bm{D}}_{0}(\omega;\bm{r},\bm{r}_{1}) \!\cdot \! \bar{\bm{D}}(\omega;\bm{r}_{1},\bm{r}').
\nonumber
\end{align}
Keeping  only the resonance (second) term and replacing   $\bar{\bm{D}}(\omega;\bm{r}_{1},\bm{r}')$  with the plasmon Green's function (\ref{dyadic-mode}), we integrate the energy flux $S=(c/8\pi)\left |\bm{\mathcal{E}}_{sc}(\omega,\bm{r})\right |^{2}$ over remote surface enclosing the  system. Using far-field asymptotics    $ \bar{\bm{D}}_{0}(\omega;\bm{r})\sim (\omega/c)^{2}(e^{ikr}/r)\left (\bm{I}-\hat{\bm{r}}\hat{\bm{r}}\right )$, we  obtain 
\begin{equation}
P_{\rm sc}(\omega)=W_{m}^{r}(\omega)|S_{m}(\omega)|^{2} [\bm{\mathcal{P}}_{m}(\omega)\!\cdot\!\bm{\mathcal{E}}_{i}]^{2}, 
\end{equation}
where  $W_{m}^{r}(\omega)$ is given by Eq.~(\ref{mode-power-rad}). Normalizing $P_{\rm sc}(\omega)$ by $S_{i}$, we obtain the mode \textit{scattering} cross section  
\begin{equation}
\label{cross-sc}
\sigma_{\rm sc}^{(m)}(\omega)=\dfrac{4\pi\omega}{c}\dfrac{\omega_{m}^{2}}{2U_{m}}\dfrac{\omega\gamma_{m}^{r}(\omega)\,[\bm{e}\!\cdot\!\bm{\mathcal{P}}_{m}(\omega)]^{2}}{(\omega_{m}^{2}-\omega^{2})^{2}+\omega^{2}\gamma_{m}^{2}(\omega)},
\end{equation}
where the plasmon radiative decay rate $\gamma_{m}^{r}(\omega)$  is given by Eq.~(\ref{mode-decay-bos-rad}). Adding $\sigma_{\rm sc}^{(m)}(\omega)$ and $\sigma_{\rm abs}^{(m)}(\omega)$ together, we obtain the mode \emph{extinction} cross section as
\begin{equation}
\label{cross-ext}
\sigma_{\rm ext}^{(m)}(\omega)=\dfrac{4\pi\omega}{c}\dfrac{\omega_{m}^{2}}{2U_{m}}\dfrac{\omega\gamma_{m}(\omega)\,[\bm{e}\!\cdot\!\bm{\mathcal{P}}_{m}(\omega)]^{2}}{(\omega_{m}^{2}-\omega^{2})^{2}+\omega^{2}\gamma_{m}^{2}(\omega)},
\end{equation}
where we used the relation $\gamma_{m}(\omega)=\gamma_{m}^{nr}(\omega)+\gamma_{m}^{r}(\omega)$, which, in this case,  implies energy flux conservation:
\begin{equation}
\label{flux-conserv}
\sigma_{\rm abs}^{(m)}(\omega) =\dfrac{\gamma_{m}^{nr}(\omega) }{\gamma_{m}(\omega) }\,\sigma_{\rm ext} ^{(m)}(\omega),
~
\sigma_{\rm sc}^{(m)}(\omega) =\dfrac{\gamma_{m}^{r}(\omega) }{\gamma_{m}(\omega) }\,\sigma_{\rm ext}^{(m)}(\omega).
\end{equation}
The full cross sections $\sigma_{abs}$, $\sigma_{sc}$ and $\sigma_{ext}$ are obtained by summing up Eqs.~(\ref{cross-abs}), (\ref{cross-sc}) and (\ref{cross-ext}) over all modes.

\subsubsection{Optical polarizability of plasmonic system}

We can now obtain optical response functions of plasmonic system by using the standard relation
\begin{equation}
\label{mode-extinct}
\sigma_{\rm ext}(\omega)=\dfrac{4\pi\omega}{c}\text{Im} [ \bm{e}\!\cdot\!\bar{\bm{\alpha}}(\omega)\!\cdot\! \bm{e}],
\end{equation}
where $\bar{\bm{\alpha}}(\omega)=\sum_{m}\bar{\bm{\alpha}}_{m}(\omega) $ is optical polarizability dyadic, which characterizes the plasmonic system's response to an external field. From Eq.~(\ref{cross-ext}), the plasmon mode polarizability is obtained explicitly as
\begin{equation}
\label{mode-polar}
\bar{\bm{\alpha}}_{m}(\omega) 
%=S_{m}(\omega)\bm{\mathcal{P}}_{m}(\omega)\bm{\mathcal{P}}_{m}(\omega)
=\dfrac{\omega_{m}^{2}}{2U_{m}}\dfrac{\bm{\mathcal{P}}_{m}(\omega) \bm{\mathcal{P}}_{m}(\omega)}{\omega_{m}^{2}-\omega^{2}-i\omega\gamma_{m}(\omega)}.
\end{equation}
The mode polarizability (\ref{mode-polar}) can be split into scattering and absorbing parts as (suppressing the $\omega$ dependence)
\begin{equation}
\label{mode-polar-sum}
\bar{\bm{\alpha}}''_{m} = \dfrac{\gamma_{m}^{r}}{\gamma_{m}} \,\bar{\bm{\alpha}}''_{m} +\dfrac{\gamma_{m}^{nr}}{\gamma_{m}}\, \bar{\bm{\alpha}}''_{m},
\end{equation}
where the first term represents the scattering contribution and satisfies the relation
\begin{equation}
\label{mode-polar-sc}
\dfrac{\gamma_{m}^{r}}{\gamma_{m}} \,\bar{\bm{\alpha}}''_{m}=\dfrac{2}{3}\left (\dfrac{\omega}{c}\right )^{3} \bar{\bm{\alpha}}_{m}\!\cdot\! \bar{\bm{\alpha}}^{*}_{m}.
\end{equation}
Since $\bar{\bm{\alpha}}_{m}$ is proportional to  the plasmonic system's volume,  the scattering is suppressed for small systems. In this case,  the  extinction is dominated by the absorption, which is given by the second term in Eq.~(\ref{mode-polar-sum}). Near the resonance, the mode polarizability takes the form
\begin{equation}
\label{mode-polar-res}
\bar{\bm{\alpha}}_{m}(\omega) 
=\dfrac{\omega_{m}}{4U_{m}}\dfrac{\bm{\mathcal{P}}_{m} \bm{\mathcal{P}}_{m}}{\omega_{m}-\omega-i\gamma_{m}/2},
\end{equation}
and, after summing over all modes, can be used to characterize the linear response of any plasmonic system supporting well-defined plasmon modes. 

The radiative decay contribution into full   polarizability, $\alpha(\omega)={\rm Tr} [\bar{\bm{\alpha}}(\omega)]$, can be expressed in general form in terms of quasistatic polarizabilities $\tilde{\alpha}_{m}(\omega)$. Taking the trace of Eq.~(\ref{mode-polar}), $\alpha(\omega)$ can be written as 

\begin{equation}
\label{polar-full}
 \alpha_{m}(\omega)=\frac{\tilde{\alpha}_{m}(\omega)}{1-i\frac{2\omega^{3}}{3c^{3}}\tilde{\alpha}_{m}(\omega)},
\end{equation}
where  
\begin{equation}
\label{mode-polar-nr}
\tilde{\alpha}_{m}(\omega) 
=\dfrac{\omega_{m}^{2}}{2U_{m}}\dfrac{\bm{\mathcal{P}}_{m}^{2}(\omega)}{\omega_{m}^{2}-\omega^{2}-i\omega\gamma_{m}^{nr}(\omega)} 
\end{equation}
is plasmon polarizability \textit{without} radiative decay. The relation (\ref{polar-full}) is similar to that for the dipole polarizability of spherical particles \cite{carminati-oc06} but, in fact, holds for any nanoplasmonic system. In a similar manner, $\alpha_{m}(\omega)$ can be shown to satisfy the optical theorem 
\begin{equation}
\label{polar-optical}
\alpha''_{m}(\omega)=\dfrac{2}{3}\left (\dfrac{\omega}{c}\right )^{3}|\alpha_{m}(\omega)|^{2}
+\frac{\tilde{\alpha}''_{m}(\omega)}{\left |1-i\frac{2\omega^{3}}{3c^{3}}\tilde{\alpha}_{m}(\omega)\right |^{2}},
\end{equation}
where the first and second terms on the right hand side describe, respectively, scattering and absorption. 

For a nanosphere with $\tilde{\alpha}_{m}(\omega)=a^{3}[\varepsilon(\omega)-1]/[\varepsilon(\omega)+2]$, by expanding $\varepsilon(\omega)$ near $\omega_{sp}$, we obtain from Eq.~(\ref{polar-full})
\begin{equation}
\label{polar-sp}
\alpha_{sp}(\omega)=\frac{3a^{3}}{\partial \varepsilon'( \omega_{sp})/\partial \omega_{sp}}\,
\dfrac{1}{\omega_{sp}-\omega-i\gamma_{sp}/2},
\end{equation}
where $\gamma_{sp}=\gamma_{sp}^{nr}+\gamma_{sp}^{r}$ is the plasmon  full decay rate with nonradiative and radiative contributions given by Eqs.~(\ref{mode-rate-nr}) and (\ref{mode-rate-rad-sp}), respectively. The same result is obtained directly from Eq.~(\ref{mode-polar-res})  (see appendix).

The approach developed in this section will be used in the rest of this paper to describe spontaneous decay of a QE coupled to plasmonic resonator.

\section{Plasmon local density of states, mode density, and mode volume  }
\label{sec4}

We are now in position to derive the plasmon LDOS  that accounts for both Ohmic  and radiative losses. On a  length scale  below the diffraction limit, surface plasmons are mostly electronic excitations interacting weakly with the radiation field. In this section we show that, within our approach, the plasmon mode volume can be defined in a natural way as the inverse of the plasmon mode density, which describes plasmon mode confinement in a local region. We derive  an explicit expression for the plasmon  mode volume at a hot spot near a sharp metal  tip and show that it scales with the metal volume while being highly sensitive to the distance from the tip.

\subsection{Mode volume for plasmonic systems}

The standard expression for the electromagnetic LDOS, $\rho(\omega,\bm{r})=(2\omega/\pi c^{2})\,\text{Im} \text{Tr} [\bar{\bm{G}}(\omega;\bm{r},\bm{r})]$, can be  written in terms of the rescaled Green dyadic (\ref{dyadic-nearfield}) as 
\begin{equation}
\rho (\omega,\bm{r})
=\dfrac{1}{2\pi^{2} \omega}\, \text{Im}\,\text{Tr}\, \bar{\bm{D}}(\omega;\bm{r},\bm{r}).
\end{equation}
Near the plasmon resonance, by using the plasmon Green dyadic (\ref{dyadic-mode-res}), we obtain the plasmon LDOS as
\begin{equation}
\label{ldos-mode}
\rho_{m}(\omega,\bm{r})
%=\frac{-1}{2\pi^{2} \omega} \, \text{Im}\,\text{Tr}\, \bar{\bm{D}}_{l}(\omega;\bm{r},\bm{r})
=\frac{1}{4\pi^{2}  W_{m}} \,\frac{ \bm{E}_{m}^{2}(\bm{r})}{1+4Q_{m}^{2}(\omega/\omega_{m}-1)^{2}},
\end{equation}
where the plasmon quality factor is given by
\begin{equation}
\label{quality}
Q_{m}=\dfrac{\omega_{m} }{\gamma_{m}}=\dfrac{\omega_{m}U_{m}}{W_{m}},
\end{equation}
and dissipated power $W_{m}=W_{m}^{nr}+W_{m}^{r}$ incorporates all plasmon damping channels. As a function of frequency, the LDOS has a Lorentzian shape and, at   resonance, is proportional to the plasmon field intensity normalized by the dissipated power \cite{shahbazyan-prl16}: $\rho (\omega_{m},\bm{r})= \bm{E}_{m}^{2}(\bm{r})/4\pi^{2}  W_{m} $.

The plasmon LDOS (\ref{ldos-mode}) describes  the plasmon states' distribution  in a unit volume and frequency interval.  Frequency integration of the LDOS  yields the \textit{plasmon mode density} 
\begin{equation}
\label{mode-density}
\rho_{m} (\bm{r})=\!\int\! d\omega \rho_{m}(\omega,\bm{r})
=\frac{\omega_{m}\bm{E}_{m}^{2}(\bm{r})}{8\pi Q_{m}W_{m} }
=\frac{\bm{E}_{m}^{2}(\bm{r})}{8\pi U_{m} },
\end{equation}
which describes spatial distribution of the plasmon field intensity. Note that, in contrast to the LDOS, $\rho (\bm{r})$  is normalized by mode energy, rather than dissipated power, and, thus, is \textit{independent of losses}. With help of Eq.~(\ref{energy-mode}), the mode density is explicitly obtained  as
\begin{equation}
\label{mode-density2}
\rho_{m} (\bm{r})
=\frac{1}{{\cal V}_{m}(\bm{r})}
= \frac{2\bm{E}_{m}^{2}(\bm{r})}{\int \! dV \bm{E}_{m}^{2}(\bm{r})\partial (\omega_{m}\varepsilon')/\partial \omega_{m}},
\end{equation}
and can be viewed as the inverse \textit{local mode volume} ${\cal V}_{m}(\bm{r})$, which characterizes the field confinement at point $\bm{r}$.  The expression (\ref{mode-density2}) is valid for any nanoplasmonic system, including plasmonic cavities and open systems.  

Note that the form (\ref{mode-density2}) for plasmon mode volume was proposed previously in the case of spherical metal nanoshell \cite{andreani-prb12}. For more general  systems described by dispersive dielectric function, a similar expression was obtained  by using expansion of full Maxwell equations' solution over quasinormal modes (QNM) \cite {lalanne-prl13}. Since QNMs are leaky modes described by complex-valued fields, the QNM volume is complex as well, and so the QNM Purcell factor is given by the real part of Eq.~(\ref{purcell}) \cite{lalanne-prl13,lalanne-lpr18}.  

Within our approach, the local mode volume at  point $\bm{r}$ arises as the inverse of the plasmon mode density at that point  and, thus, represents a \textit{real} function of plasmon field intensity and is independent of radiative and nonradiative losses. These losses still affect the Purcell factor (\ref{purcell}) because they determine the quality factor $Q_{m}$  via the full plasmon decay rate $\gamma_{m}=\gamma_{m}^{r}+\gamma_{m}^{nr}$, thereby ensuring energy flux conservation.

\subsection{Plasmon mode volume near metallic tip }

The largest plasmonic enhancements occur if   QE is located at a  \textit{hot spot}---a small region characterized by very high mode density (or very small mode volume), e.g., near a sharp tip of a metal nanostructure. With help of Eq.~(\ref{mode-density2}),  the maximal mode density  can be estimated  by assuming the classical field profile near the metal surface. Due to Gauss's law, the local fields do  not significantly change  inside the small metallic structure, while falling off rapidly  outside of it, so the highest field intensity is achieved near the metal surface,
\begin{equation}
\rho_{m}(\bm{r})\approx  \frac{2}{\omega_{m}\partial \varepsilon'(\omega_{m})/\partial \omega_{m}}\frac{ E_{L}^{2}(\bm{r})+E_{T}^{2}(\bm{r})  }{V_{\rm met} \left ([E_{L}^{in}]^{2}+E_{T}^{2}\right )},
\end{equation}
where  $V_{\rm met}$ is the metal volume. Here, subscripts $L$ and $T$ stand for longitudinal (normal to the tip) and transverse (tangential to the tip) field components, and superscripts $in$ and $out$ indicate local fields at the interface on the metal and dielectric sides, respectively.  The highest field localization is achieved when $E_{T}$, which is continuous across the metal-dielectric interface, is much smaller than $E_{L}$. Assuming that the local field is polarized  along the tip, i.e., $E_{L}\gg E_{T}$, and using the boundary condition for the normal  field component $E_{L}^{out}=\varepsilon'(\omega_{m})E_{L}^{in}$, we obtain the mode density projected along the tip:
\begin{equation}
\label{mode-volume-tip}
\rho_{L}(\bm{r}) =\dfrac{1}{ {\cal V}_{L}(\bm{r})} = \dfrac{1}{V_{\rm met}}\,\frac{2|\varepsilon'(\omega_{m})|^{2} \,\tilde{E}_{L}^{2}(\bm{r})}{\omega_{m} \partial \varepsilon'(\omega_{m})/\partial \omega_{m}} ,
\end{equation}
where  $\tilde{E}_{L}(\bm{r})=  E_{L}(\bm{r})/E_{L}^{out}$ is the normal field component at point $\bm{r}$ near the tip normalized by its value at the tip. Although  the mode volume near a hot spot scales with the  metal volume $V_{\rm met}$,   the ratio $V_{\rm met}/{\cal V}_{L}=V_{\rm met}\rho_{L}$ depends  on  the proximity  of QE to the tip. While the mode density is highest at the tip ($\tilde{E}_{L}=1$),  it is expected to saturate  below distances $\sim v_{F}/\omega$ because the nonlocal effects become dominant \cite{mortensen-nc14,mortensen-optica17}. Note that, for noble metals,  this length scale is $\sim 1$ nm in the plasmonic frequency range.

\section{Purcell factor and enhancement factor for power spectrum}
\label{sec5}

Purcell factor characterizes  the enhancement of  QE decay rate due to ET between QE and the plasmonic resonator.  Part of the transferred energy  is radiated away by the plasmonic antenna, while the rest is  dissipated due to the Ohmic losses in metal. In this section, we derive explicit expressions for the Purcell factor for spontaneous decay rate and   the enhancement factor for the radiated power spectrum. In this paper, we only consider the weak-coupling regime and disregard plasmon back action on the QE spectrum.

\subsection{Quantum-emitter-plasmon energy-transfer rate and Purcell factor}

The ET rate  between a QE situated at $\bm{r}_{0}$ with dipole moment $ \bm{p}= \mu\bm{n}$ and a resonant plasmon mode is straightforwardly obtained from Eq.~(\ref{rate-et}) by using the plasmon Green's function  (\ref{dyadic-mode-res}) as 
\begin{equation}
\label{rate-mode-u}
\gamma_{et}(\omega)
%=-\frac{\mu^{2}}{\hbar}\,  \bm{n}\!\cdot\!\bar{\bm{D}}''(\omega_{m};\bm{r},\bm{r}_{0})\!\cdot \!\bm{n} 
=\frac{\mu^{2}Q_{m}}{\hbar U_{m}}\, 
\frac{[\bm{n}\!\cdot\!\bm{E}_{m}(\bm{r}_{0})]^{2}}{1+4Q_{m}^{2}(\omega/\omega_{m}-1)^{2}}.
\end{equation}
As a function of  the QE emission frequency $\omega$, the  rate (\ref{rate-mode-u}) has a Lorentzian shape with  maximum at  $\omega=\omega_{m}$.  In terms of mode volume projected  on the QE dipole direction $\bm{n}$,
\begin{equation}
\label{mode-density-proj}
\rho_{m}^{n}(\bm{r})=\dfrac{1}{{\cal V}_{m}^{n}(\bm{r})} =\frac{2\left [\bm{n}\!\cdot\!\bm{E}_{m}(\bm{r})\right ]^{2}}{\int \! dV \bm{E}_{m}^{2}\partial (\omega_{m}\varepsilon')/\partial \omega_{m}}, 
\end{equation}
the QE-plasmon ET rate  takes the form
\begin{equation}
\label{rate-mode}
\gamma_{et}(\omega)
%=-\frac{\mu^{2}}{\hbar}\,  \bm{n}\!\cdot\!\bar{\bm{D}}''(\omega_{m};\bm{r},\bm{r})\!\cdot \!\bm{n} 
=\frac{8\pi\mu^{2}}{\hbar {\cal V}_{m}^{n}(\bm{r}_{0})}\, 
\frac{Q_{m}}{1+4Q_{m}^{2}(\omega/\omega_{m}-1)^{2}}.
\end{equation}
Normalizing the QE-plasmon ET rate at the resonance frequency, $\gamma_{et}(\omega_{m}) =8\pi\mu^{2}Q_{m}/\hbar{\cal V}_{m}^{n}$, by the free-space QE spontaneous decay rate (\ref{rate-free}), we finally obtain the Purcell factor for a QE coupled to resonant plasmon mode, 
\begin{equation}
\label{purcell-plas}
F_{p}=\frac{6\pi Q_{m}}{k^{3}{\cal V}_{m}^{n}}= \frac{12\pi Q_{m}\left [\bm{n}\!\cdot\!\bm{E}_{m}(\bm{r}_{0})\right ]^{2}}{k^{3}\! \int \! dV \bm{E}_{m}^{2}\partial (\omega_{m}\varepsilon')/\partial \omega_{m}},
\end{equation}
which extends the cavity Purcell factor (\ref{purcell})  to plasmonic resonators. For a QE   at the hot spot near a metallic tip, with help of Eq.~(\ref{mode-volume-tip}), we obtain
\begin{equation}
\label{purcell-plas-tip}
F_{p}^{\rm tip}=
\frac{12\pi Q_{m}|\varepsilon'(\omega_{m})|^{2}}
{k^{3} V_{\rm met}\omega_{m}\partial \varepsilon'(\omega_{m})/\partial \omega_{m}} 
\bigl[ \bm{n}\!\cdot\!\tilde{\bm{E}}_{L}(\bm{r}_{0})\bigr]^{2},
\end{equation}
where $\bm{n}\!\cdot\!\tilde{\bm{E}}_{L}(\bm{r}_{0})$ stands for the projection of the normalized field component  along the tip onto the QE's dipole orientation $\bm{n}$. The Purcell factor is maximal when the QE dipole is oriented along the tip  whereas, for transverse dipole orientation, there is no significant enhancement.

\subsection{Radiated power spectrum }

Part of the energy transferred from the  QE to the resonant plasmon mode is radiated away by the plasmonic antenna, leading to an overall enhancement of the radiated power   observed, e.g., in plasmon-enhanced fluorescence experiments \cite{feldmann-prl02,artemyev-nl02,novotny-prl06,sandoghdar-prl06,halas-nl07,halas-acsnano09,ming-nl09}.  While a plasmon's radiative decay rate (\ref{mode-rate-rad}) is typically much larger than that of individual QEs, i.e., $\gamma_{m}^{r}\gg \gamma_{0}^{r}$, a significant part of the transferred energy is dissipated in  the metal  at rate (\ref{mode-rate-nr}), so that the enhancement factor depends on the radiation efficiency of the plasmonic antenna $\eta_{m}=\gamma_{m}^{r}/\gamma_{m}$.

The power radiated by a QE  placed at position $\bm{r}_{0}$ near a plasmonic antenna is obtained by integrating Poynting's vector $S=(c/8\pi)|\bm{\mathcal{E}}(\bm{r})|^{2}$ over a remote surface enclosing the   system, where $\bm{\mathcal{E}}(\bm{r})$ is the QE electric field  \cite{novotny-book}: 
\begin{align}
\label{electric-qe-single}
\bm{\mathcal{E}}(\bm{r} ) =  \bar{\bm{D}}(\omega;\bm{r},\bm{r}_{0}) \!\cdot \!\bm{p}, 
\end{align}
and $\bar{\bm{D}}(\omega;\bm{r},\bm{r}_{0})$ is the Green dyadic (\ref{dyadic-nearfield}). To extract  the far field contribution, we  use the Dyson equation  (\ref{dyson}). Replacing the near-field Green dyadic $\bar{\bm{D}}$ in the integrand by the plasmon Green dyadic (\ref{dyadic-mode-res}), the QE-generated far field (\ref{electric-qe-single}) takes the form 
\begin{align}
\label{electric-qe-single2}
\bm{\mathcal{E}}(\bm{r}) 
&=   \bar{\bm{D}}_{0}(\omega;\bm{r}-\bm{r}_{0}) \!\cdot \!\bm{p}\\
+&\frac{\omega_{m}}{4 U_{m}}\frac{\bm{E}_{m}(\bm{r}_{0})\!\cdot\!\bm{p}}{\omega_{m}-\omega-i\gamma_{m}/2}\int \! dV' \bar{\bm{D}}_{0}(\omega;\bm{r}-\bm{r}') \!\cdot\!\bm{P}_{m}(\bm{r}').
\nonumber
\end{align}
Straightforward integration of Poynting's vector over remote spherical surface yields the radiated power
\begin{equation}
\label{mode-power-rad1}
W_{r}(\omega)= \frac{\omega^{4}}{3c^{3}} \left | \bm{p}+\frac{\omega_{m}}{4 U_{m}}\frac{\bm{\mathcal{P}}_{m}\, [\bm{E}_{m}(\bm{r}_{0})\!\cdot\!\bm{p}]}{\omega_{m}-\omega-i\gamma_{m}/2}\right |^{2},
\end{equation}
where the second term represents the contribution of the plasmonic antenna with dipole moment $\bm{\mathcal{P}}_{m}$. Near the resonance, the plasmon emission is dominant  and, disregarding the first nonresonant term,  we obtain
\begin{equation}
\label{mode-power-rad2}
W_{r}(\omega)
=\frac{\mu^{2}\omega^{4}}{3c^{3}}
\frac{\gamma_{m}^{r}\gamma_{et}(\omega)}{\gamma_{m}\gamma_{0}^{r}},
\end{equation}
where the QE-plasmon ET rate $\gamma_{et}(\omega)$ is given by Eq.~(\ref{rate-mode}), and radiative decay rates $\gamma_{0}^{r}$ and $\gamma_{m}^{r}$  are given by Eqs.~(\ref{rate-free}) and (\ref{mode-rate-rad}), respectively. Normalizing $W_{r}(\omega)$  by the spectral power $W_{r}^{0}=\mu^{2}\omega^{4}/3c^{3}$ radiated by an isolated QE \cite{novotny-book}, we obtain  the enhancement factor for the power spectrum
\begin{equation}
\label{mode-power-rad3}
M(\omega)=
%\frac{\gamma_{m}^{r}\gamma_{et}(\omega)}{\gamma_{m}\gamma_{0}^{r}}=
\frac{F_{p}\eta_{m}}{1+4Q_{m}^{2}(\omega/\omega_{m}-1)^{2}},
\end{equation}
where the  Purcell factor $F_{p}$ is given by Eq.~(\ref{purcell-plas}) and the plasmon radiation efficiency $\eta_{m}$ is given by (\ref{mode-efficiency}). At the  resonance, $|\omega-\omega_{m}|\ll \gamma_{m}$, we obtain
\begin{equation}
\label{enhance}
M(\omega_{m})=F_{p}\eta_{m}=\frac{6\pi Q_{m}}{k^{3}{\cal V}_{m}^{n}}\, \eta_{m},
\end{equation}
which represents the general relation between the Purcell factor for spontaneous decay and the maximal enhancement factor.  For high radiation efficiency $\eta\sim 1$, the enhancement factor is comparable to the Purcell factor, i.e., energy is radiated by the plasmonic antenna at approximately the same rate as it is being received from the QE.

Note finally that the relation (\ref{enhance}) overestimates the enhancement factor because it does not account for   ET from the QE to off-resonant modes which leads to radiation quenching at close QE-metal distances. The fraction of energy  transferred to a bright plasmon mode is $q=F_{p}/\sum_{l}F_{p}^{(l)}$, where $F_{p}^{(l)}$ are Purcell factors for all modes  and so, close to the metal surface, the enhancement factor $M$ is suppressed by the quenching factor $q$.

\section{Numerical results and discussion}
\label{sec6}

\begin{figure}[tb]
%\centering
\begin{center}
\includegraphics[width=0.99\columnwidth]{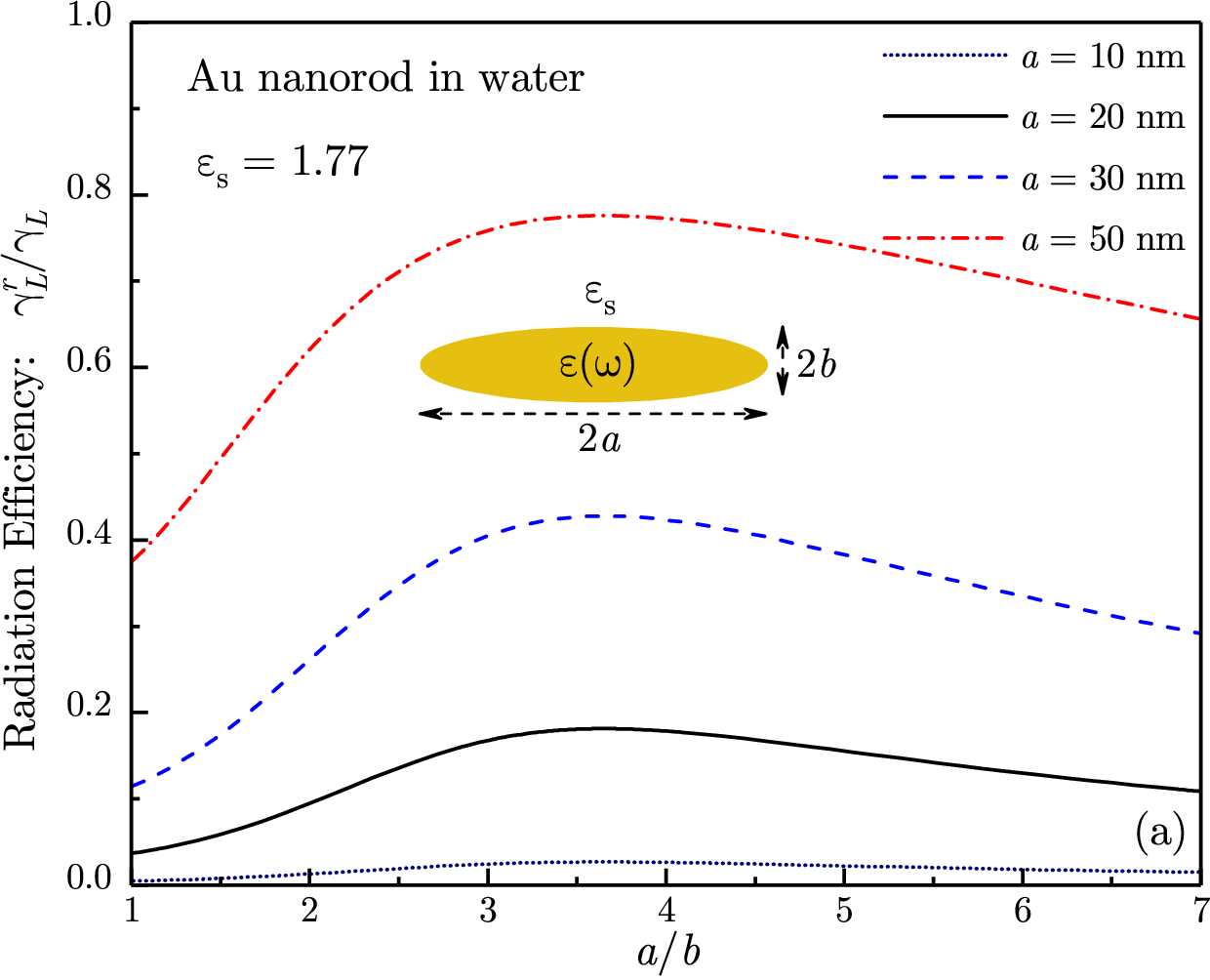}
\\
\vspace{2mm}
\includegraphics[width=0.99\columnwidth]{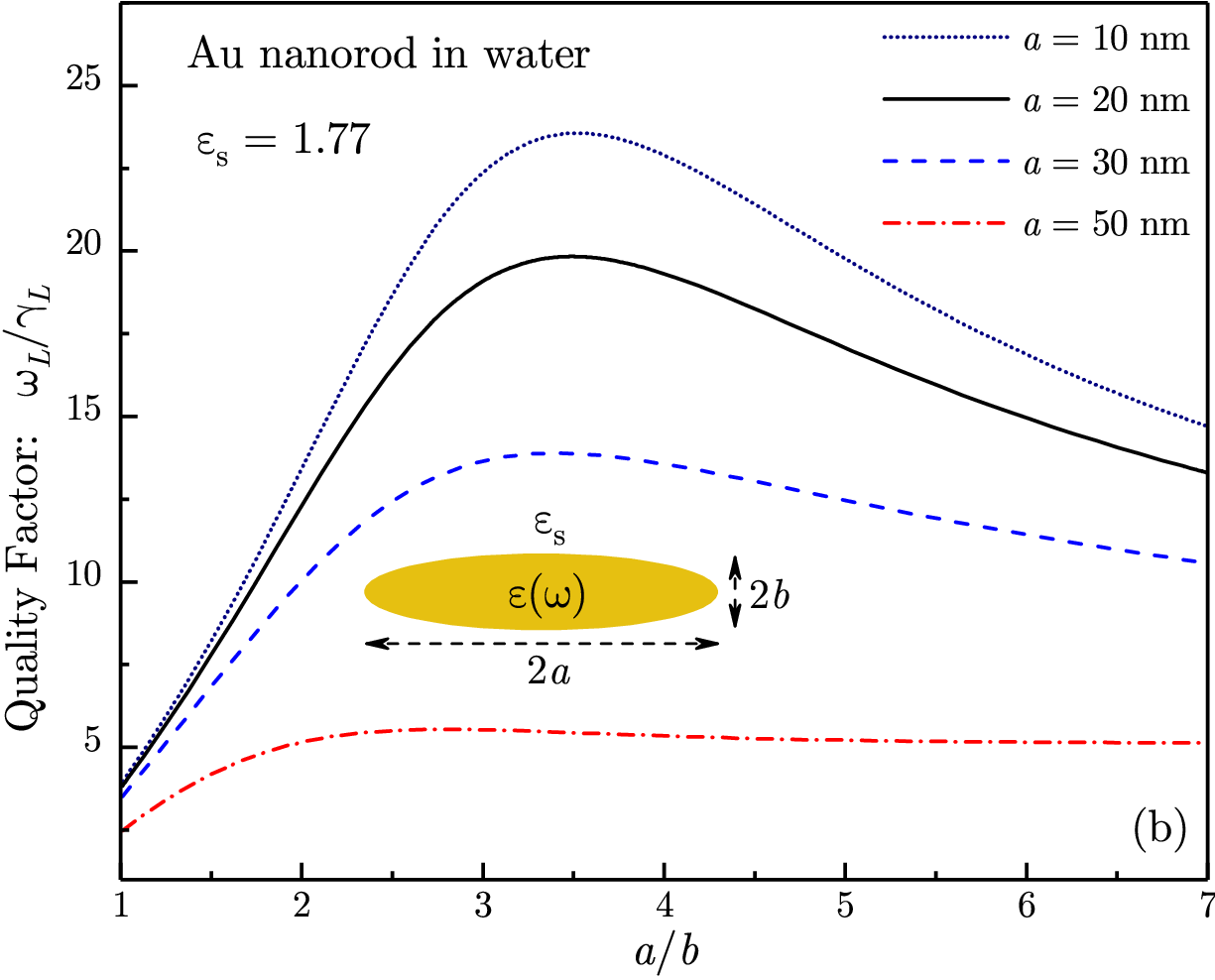}
\caption{\label{fig1}  
Plasmon radiation efficiency $\eta_{L}$ (a), and quality factor $Q_{L}$ (b)   plotted against aspect ratio $a/b$ for different  nanorod sizes. Insets show schematics of a  prolate spheroidal particle.}
\end{center}
\vspace{-8mm}
\end{figure}

To illustrate our theory, we performed numerical calculations for a QE coupled to longitudinal  plasmon mode oscillating, with frequency $\omega_{L}$, along a Au nanorod, which is modeled here by prolate spheroid with semimajor and semiminor axes $a$ and $b$, respectively (see schematics in Fig.~\ref{fig1}).  This needle-shaped   structure  is characterized by a relatively high radiation efficiency  while, at the same time, it possesses hot spots near the tips, where the plasmon field is  highly localized. We assume that the Au nanorod is submerged in water ($\varepsilon_{s}=1.77$) and use the experimental Au dielectric function $\varepsilon(\omega)=\varepsilon'(\omega)+i\varepsilon''(\omega)$ in all calculations. The dielectric constant $\varepsilon_{s}$  of the surrounding medium is restored in all expressions via the replacements: $c\rightarrow c/\varepsilon_{s}$, $\varepsilon(\omega,\bm{r}) \rightarrow \varepsilon(\omega,\bm{r})/\varepsilon_{s}$, and $\mu^{2}\rightarrow \mu^{2}/\varepsilon_{s}$.  Analytical  expressions for spheroidal particles  are provided in the appendix along with other technical details, and here we  only discuss the results of numerical calculations.

%
%%%%%%%%%%%%%%%%%%%%%%%%%%%%%%%%%%%%%%%%%%%%%%
%
%
%%%%%%%%%%%%%%%%%%%%%%

In Fig.~\ref{fig1}, we show the calculated plasmon radiation efficiency $\eta_{L}=\gamma_{L}^{r}/\gamma_{L}$ and quality factor $Q_{L}=\omega_{L}/\gamma_{L}$, which include both radiative and Ohmic losses. As expected, the increase of $\eta_{L}$ [see Fig. \ref{fig1}(a)] due to the increase of $\gamma_{L}^{r}$ with overall nanorod size  is accompanied by the reduction of the quality factor [see Fig. \ref{fig1}(b)] due to overall increase of the plasmon decay rate $\gamma_{L}=\gamma_{L}^{r}+\gamma_{L}^{nr}$. The maximal values of $\eta_{L}$ and $Q_{L}$ are reached for the aspect ratio $a/b$ in the range  $3- 5$, corresponding to plasmon wavelength range $650- 800$ nm. In this range, $\varepsilon''(\omega)$ for Au reaches its minimum, which translates to the lowest  Ohmic losses and, thus, the highest  $\eta_{L}$ and $Q_{L}$, except for the largest nanorod ($a=50$ nm), where the plasmon decay is dominated by the radiative channel  [see Fig. \ref{fig1}(b)].

%%%%%%%%%%%%%%%%%%%%%%%%%%%%%%%%%%%%%%%
%
\begin{figure}[tb]
%\centering
\begin{center}
\includegraphics[width=0.99\columnwidth]{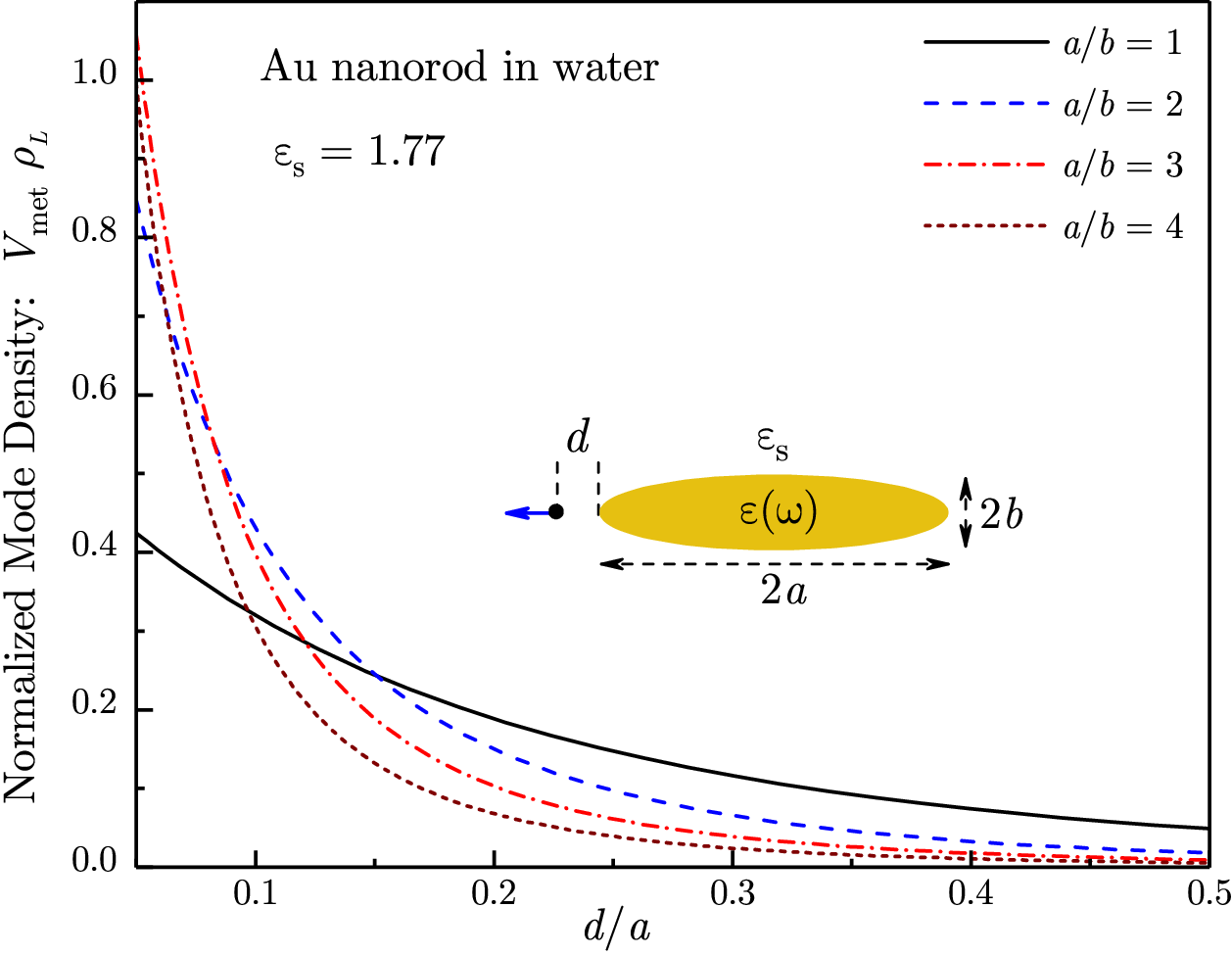}
\caption{\label{fig2}  
Normalized mode density (inverse mode volume) projected along the Au nanorod is plotted against the distance to the  nanorod tip for different aspect ratios $a/b$ at fixed   $a$.}
\end{center}
\vspace{-8mm}
\end{figure}
%
%%%%%%%%%%%%%%%%%%%%%%%%%%%%%%%%%%%%%%%%%%%%%

To study the field confinement at a hot spot,  we plot in Fig.~\ref{fig2} the projected plasmon mode density $\rho_{L}$, normalized by  the metal volume, as a function of distance $d$ to the nanorod tip for several values of aspect ratio.  Note that, for spheroidal particles,  Eq.~(\ref{mode-volume-tip}) is exact. To account for field-enhancement saturation due to  nonlocal effects \cite{mortensen-nc14,mortensen-optica17}, we restrict the minimal distance to the tip by   $d_{\rm min}=0.05a$, and change the nanorod volume by reducing $b$ at fixed $a$. For aspect ratios $a/b$ in the range $2 - 4$, i.e., when hot spots at the tips are well developed, the mode volume ${\cal V}_{L}=1/\rho_{L}$ exhibits nearly universal behavior reaching $V_{\rm met}$ in the hot spot region while rapidly decreasing when moving away from the tip.

%%%%%%%%%%%%%%%%%%%%%%%%%%%at%%%%%%%%%%%%%%%%%%%
%
\begin{figure}[tb]
%\centering
\begin{center}
\includegraphics[width=0.99\columnwidth]{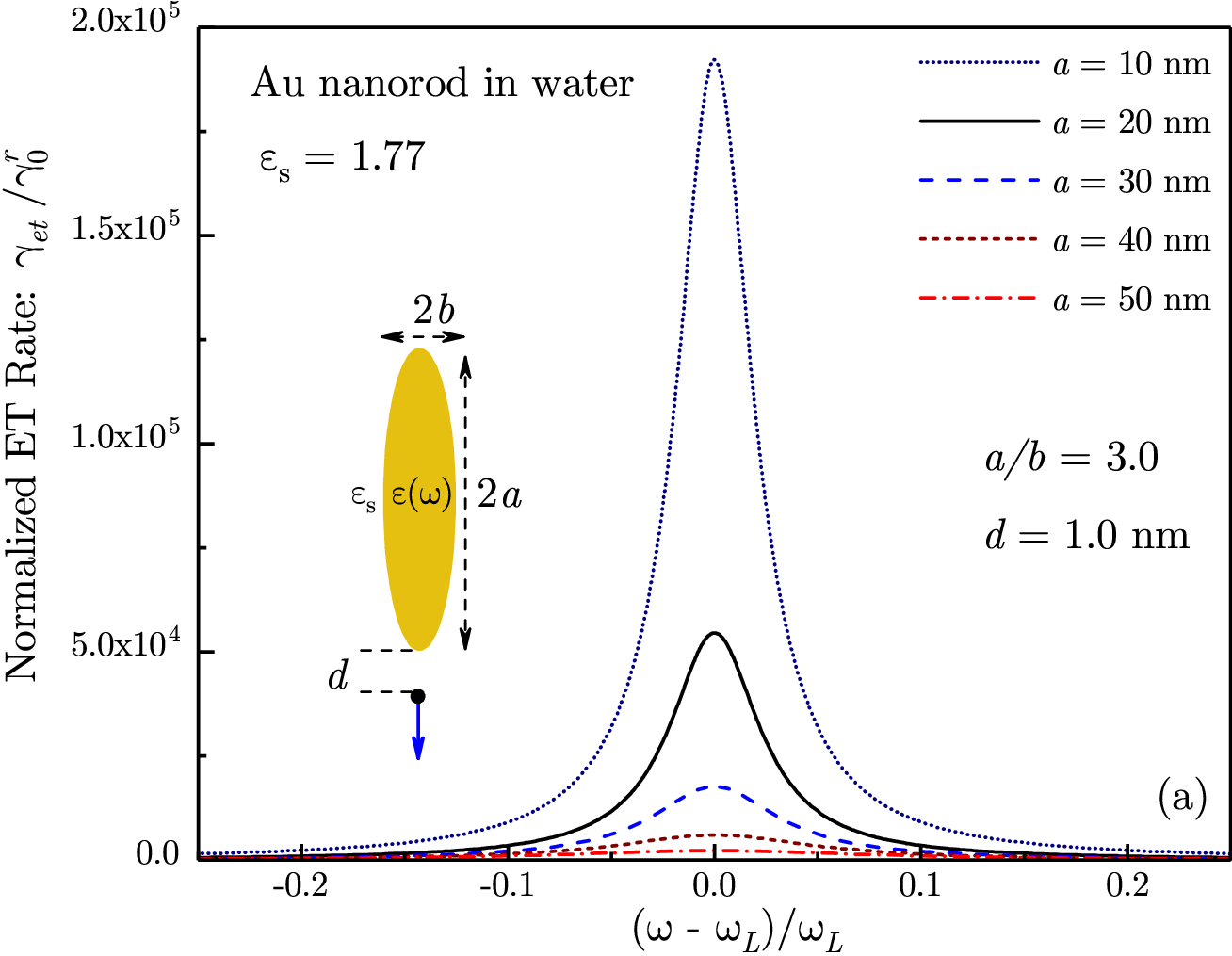}
\\
\vspace{2mm}
\includegraphics[width=0.99\columnwidth]{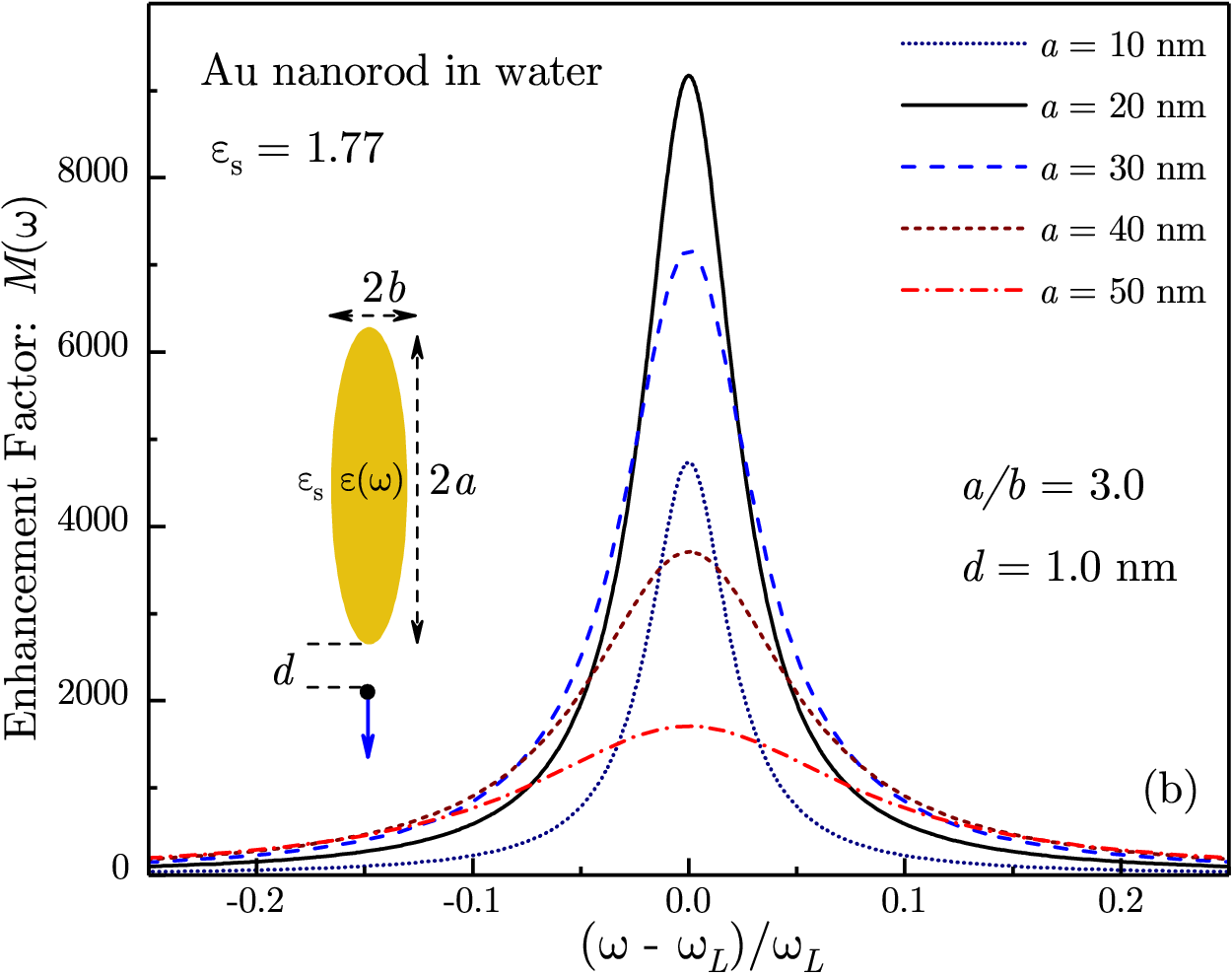}
\caption{\label{fig3}  
Frequency dependence of normalized QE-plasmon ET rate and enhancement factor for power spectrum for normally-oriented QE at a distance of 1.0 nm from Au nanorod tip is plotted for different nanorod sizes and fixed aspect ratio $a/b=3.0$.}
\end{center}
\vspace{-8mm}
\end{figure}
%
%%%%%%%%%%%%%%%%%%%%%%

Consider now spontaneous decay of a QE  at distance $d$ from the nanorod tip with its dipole oriented normally to the metal surface (see schematics in Fig.~\ref{fig3}). We assume that the QE is situated at a fixed distance  $d=1$ nm from the  tip, where the plasmon field is highly localized. In Fig.~\ref{fig3}, we show the QE-plasmon  ET rate (\ref{rate-mode}), normalized by the free-space decay rate (\ref{rate-free}), and the  enhancement factor for the power spectrum (\ref{mode-power-rad3}) plotted against  QE emission frequency $\omega$ for different overall sizes  but at fixed aspect ratio $a/b=3.0$. The amplitude of frequency Lorentzian $\gamma_{et}(\omega)/\gamma_{0}^{r}$  in Fig.~\ref{fig3}(a) is given by the Purcell factor (\ref{purcell-plas}), which, near the hot spot, scales as $Q_{L}/k^{3}V_{\rm met}$ [see Eq.~(\ref{purcell-plas-tip})]. With increasing nanorod size, the Purcell factor sharply decreases due to combined effect of decreasing $Q_{L}$ and, more importantly, increasing $k^{3}V_{\rm met}$. However, the enhancement factor  $M(\omega)$ in Fig.~\ref{fig3}(b) exhibits more complicated behavior: its amplitude $F_{p}\eta_{L}$   first sharply increases due to rapid change of $\eta_{L}$ as $a$ changes from 10 nm to 20 nm,  but, then,  for larger $a$, falls down as the metal volume effect in $F_{p}$ takes over. 

\begin{figure}[tb]
%\centering
\begin{center}
\includegraphics[width=0.99\columnwidth]{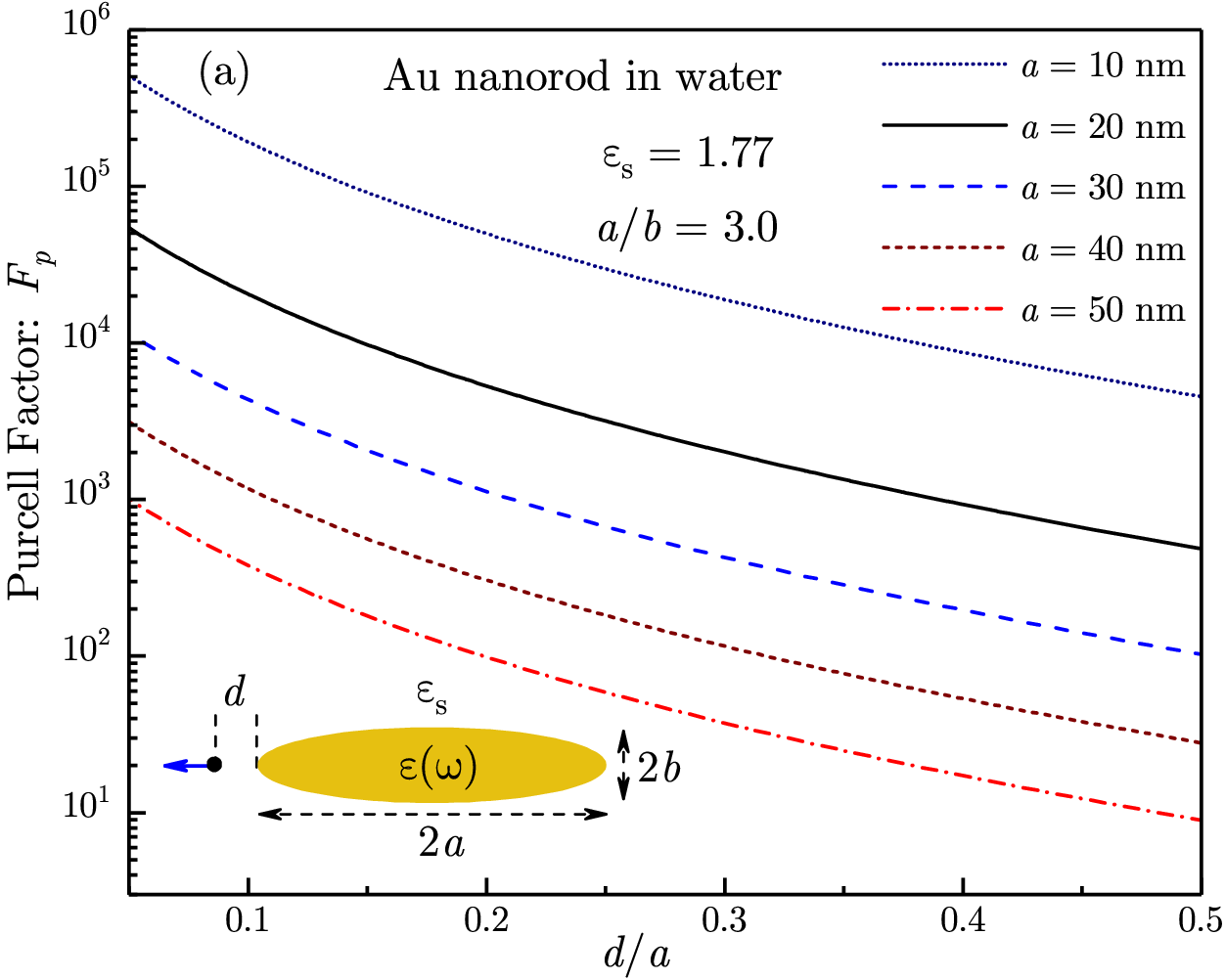}
\\
\vspace{2mm}
\includegraphics[width=0.99\columnwidth]{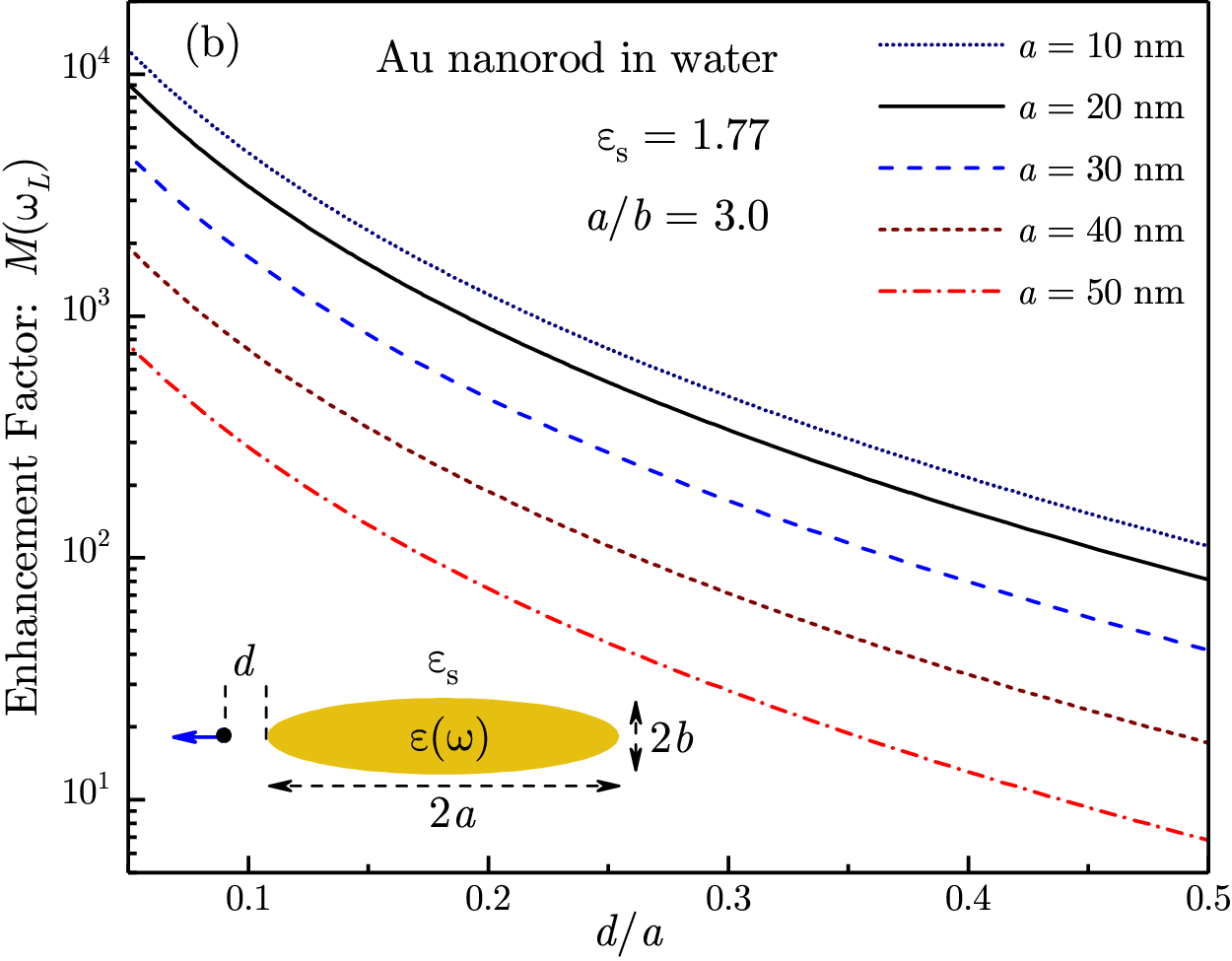}
\caption{\label{fig4}  
Distance dependence of Purcell factor and enhancement factor for power spectrum at resonance frequency is plotted for normally oriented QE for different Au nanorod sizes and fixed aspect ratio $a/b$.}
\end{center}
\vspace{-8mm}
\end{figure}

In Fig.~\ref{fig4}, we show the  Purcell factor $F_{p}$ and  enhancement factor at resonance frequency $M(\omega_{L})=F_{p}\eta_{L}$ plotted against the distance $d$ to the nanorod tip for several  overall sizes. With the QE moving away from the tip, both $F_{p}$ and $M(\omega_{L})$ decrease by up to two orders of magnitude as $d$ increases to $a/2$, indicating that the plasmon field is highly localized near the  tips (see Fig.~\ref{fig2}).   Note that, since the distance  in Fig.~\ref{fig4} is measured in units of nanorod size, the same starting point $d=0.05a$ for each curve  translates into different initial distances  to the metal surface. After appropriate rescaling to bring initial distances to  the same numerical value (e.g., 1.0 nm), the order of curves in Fig.~\ref{fig4} follows that in Fig.~\ref{fig3}. Overall, Figs.~\ref{fig3} and \ref{fig4} indicate that the Purcell factor and enhancement factor are highly sensitive to the system size due to scaling of the plasmon mode volume with the metal volume (see Fig.~\ref{fig2})  and, to lesser degree, size-dependence of  plasmon quality factor and radiation efficiency (see Fig.~\ref{fig1}).

%%%%%%%%%%%%%%%%%%%%%%%%%%%%
\section{Conclusions}
\label{sec7}

In summary, we presented herein a theory for spontaneous decay of a quantum emitter coupled to a localized plasmon mode in a  metal-dielectric structure characterized by a dispersive dielectric function which incorporates, in a consistent way, plasmon coupling to the radiation field. For plasmonic systems with characteristic size below the diffraction limit, we derived explicit expressions for plasmon radiative decay rate, which determines radiation efficiency of a plasmonic antenna, and optical polarizability, which defines system response to an external field. Using these results, we extended our approach \cite{shahbazyan-prl16} to derive plasmon Green's function that now includes plasmon interaction with radiation field  and obtained explicit expression  for the plasmon local density of states that accounts for all relevant plasmon damping channels. We have shown that plasmon mode volume is defined naturally as the inverse of plasmon mode density, which characterized plasmon field confinement, and that, for well-defined plasmon modes, it is independent of losses. We estimated the plasmon mode volume at a hot spot near a sharp tip of a small metal nanostructure and demonstrated that it scales with the metal volume, although its actual value is highly sensitive to the QE distance to the tip. Using our approach, we recovered the usual form of the Purcell factor, but now for plasmonic resonators, and established its relation  with the enhancement factor for radiated power. Finally, we illustrated our approach by presenting numerical results for QE situated near the tip of a Au nanorod.

%%%%%%%%%%%%%%%%%%%%%%%%%%%%%%%
This work was supported in part by National Science Foundation under grant No. DMR-1610427,  No. DMR-1826886, and No. HRD-1547754.

\appendix

%%%%%%%%%%%%%%%%%%%%%%%%%%%%%%%%%%%%%%%%%%%%%%%

\section{Potentials and fields in nanospheroids}

Consider a prolate spheroid with semiaxis $a$ along the symmetry axis and semiaxis  $b$ in the symmetry plane ($a>b$). We use standard notations for spheroidal coordinates ($\xi,\eta,\phi$) where $\xi$ is the "radial" coordinate while $\eta=\cos \theta$ and $\phi$ parametrize the surface.  The scaling factors are given by
\begin{align}
&h_{\xi}=f\sqrt{\frac{\xi^{2}-\eta^{2}}{\xi^{2}-1}},
~~
h_{\eta}=f\sqrt{\frac{\xi^{2}-\eta^{2}}{1-\eta^{2}}},
\nonumber\\
&h_{\phi}=f\sqrt{(\xi^{2}-1)(1-\eta^{2})},
\end{align}
where $f =\sqrt{a^{2}-b ^{2}}$ is half distance between the foci, and spheroid surface corresponds to $\xi_{1}=a/f$. The volume and surface elements are, respectively, $dV=h_{\xi}h_{\eta}h_{\phi} d\xi d\eta d\phi$ and $dS=h_{\eta}h_{\phi} d\eta d\phi$, and the gradient operator is $\bm{\nabla}=\hat{\bm{\xi} }h_{\xi}^{-1}\partial/\partial\xi+\hat{\bm{\eta} }h_{\eta}^{-1}\partial/\partial\eta+\hat{\bm{\phi} }h_{\phi}^{-1}\partial/\partial\phi$, where $\hat{\bm{\xi} }$, $\hat{\bm{\eta}}$ and $\hat{\bm{\phi} }$ are spheroidal unit vectors,
\begin{align}
\label{unit-spheroid}
&\hat{\bm{\xi} }=\frac{1}{h_{\xi}}\frac{\partial \bm{r}}{\partial \xi}=
\frac{f\xi}{h_{\eta}}
\left (\cos\phi \hat{\bm{x}} + \sin\phi \hat{\bm{y}}\right )
+ \frac{f\eta}{h_{\xi}}\hat{\bm{z}},
\nonumber\\
&\hat{\bm{\eta} }=\frac{1}{h_{\eta}}\frac{\partial \bm{r}}{\partial \eta}=
-\frac{f\eta}{h_{\xi}}
\left (\cos\phi \hat{\bm{x}} + \sin\phi \hat{\bm{y}}\right )
+ \frac{f\xi}{h_{\eta}}\hat{\bm{z}},
\nonumber\\
&\hat{\bm{\phi} }=\frac{1}{h_{\phi}}\frac{\partial \bm{r}}{\partial \phi}=
-\sin\phi \hat{\bm{x}} + \cos\phi \hat{\bm{y}}.
\end{align}
The volume and surface elements  are, respectively, $dV=h_{\xi}h_{\eta}h_{\phi} d\xi d\eta d\phi$ and $dS=h_{\eta}h_{\phi} d\eta d\phi$, while the full surface area  and volume are
\begin{equation}
S=2\pi b^{2}\left (1+\frac{2\alpha}{\sin 2\alpha}\right ),
~~
V= \frac{4\pi}{3}\, a b^{2},
\end{equation}
where $\alpha=\arccos (b/a)$ is the angular eccentricity.  For $b/a\ll 1$, we have $S=\pi^{2} ab$.

The potentials for longitudinal and transverse dipole modes are 
%$\Phi_{L}=fR_{L}(\xi)P_{1}(\eta)$ and $\Phi_{T}=fR_{T}(\xi)P_{1}^{1}(\eta)\cos\phi$. 
%
\begin{equation}
\label{potentials}
\Phi_{L}=fR_{L}(\xi)P_{1}(\eta),
~~~
\Phi_{T}=fR_{T}(\xi)P_{1}^{1}(\eta)\cos\phi.
\end{equation}
For a metallic spheroid with permittivity $\varepsilon(\omega)$ in a medium with dielectric constant $\varepsilon_{s}$, the radial components for the longitudinal mode are 
\begin{align}
\label{R-long}
&R_{L}(\xi)=P_{1}(\xi),~~~\text{for $\xi<\xi_{1}$},
\nonumber\\
&R_{L}(\xi)=Q_{1}(\xi)P_{1}(\xi_{1})/Q_{1}(\xi_{1}),~~~\text{for $\xi>\xi_{1}$}.
\end{align}
The  plasmon frequencies $\omega_{L}$   follow from the continuity of $\varepsilon R'(\xi)$ across the metal/dielectric interface.

\section{Plasmon energy in spheroidal particles}

In the quasistatic approximation, the plasmon mode energy comes solely from the metal  and has the form
\begin{align}
U_{m}
&=\frac{\omega_{m}}{16\pi}\dfrac{\partial \varepsilon(\omega_{m})}{\partial \omega_{m}}
\int \!dV_{\rm met} \bm{E}_{m}^{2}
\nonumber\\
&=\frac{\omega_{m}}{16\pi}\dfrac{\partial \varepsilon(\omega_{m})}{\partial \omega_{m}}
\int \!dS  \Phi {\nabla}_{n}\Phi, 
\end{align}
where $V_{\rm met}$ and $S $ are the volume and surface of metal nanoparticle, respectively, and ${\nabla}_{n}$ is the normal derivative. Using Eqs.~(\ref{R-long}), we obtain 
\begin{equation}
\label{mode-energy-spheroid}
U_{m}=V_{\rm met}\,\frac{\omega_{m}}{16\pi}\dfrac{\partial \varepsilon(\omega_{m})}{\partial \omega_{m}}
=\,ab^{2}\,\frac{\omega_{m}}{12}\dfrac{\partial \varepsilon(\omega_{m})}{\partial \omega_{m}}.
\end{equation}

\section{Plasmon radiative decay in spheroidal particles}

The  decay rate  of a plasmon mode in metal-dielectric system  has the form
\begin{equation}
\gamma_{m}^{r}=\frac{\omega_{m}^{4}}{3c^{3}} \frac{\bm{\mathcal{P}}_{m}^{2}}{U_{m}}
\end{equation}
where $\bm{\mathcal{P}}_{m} =\chi'(\omega_{m})\int \! dV_{\rm met} \bm{E}_{m}(\bm{r})$ is the plasmon dipole moment. Due to Gauss's law, $\bm{\mathcal{P}}_{m}$ can be written as the  surface integral
\begin{align}
\bm{\mathcal{P}}_{m} 
&=-\chi'(\omega_{m})\int\! dS \Phi_{m}\hat{\bm{\xi}}
\nonumber\\
&=-2\pi \chi'(\omega_{m}) f^{2} \xi_{1}\int_{-1}^{1} d\eta \eta^{2}\,\frac{h_{\phi}h_{\eta}}{h_{\xi}} \hat{\bm{z}},
\end{align}
where we used  Eqs.~(\ref{unit-spheroid}) and (\ref{potentials}), yielding 
\begin{equation}
\label{mode-dipole-spheroid-squared}
 \bm{\mathcal{P}}_{m} =-\chi'(\omega_{m})\frac{4\pi}{3}f^{3}\xi_{1}(\xi_{1}^{2}-1)\hat{\bm{z}}=-\chi'(\omega_{m}) V_{\rm met}\hat{\bm{z}}.
\end{equation}
Finally, using Eqs.~(\ref{mode-dipole-spheroid-squared}) and (\ref{mode-energy-spheroid}), the plasmon radiative decay rate is evaluated as 
\begin{equation}
\gamma_{L}^{r}=\dfrac{\omega_{m}^{3}\sqrt{\varepsilon_{s}}}{3\pi c^{3}}\frac{\left [\varepsilon'(\omega_{L})-\varepsilon_{s}\right ]^{2}V_{\rm met}}{\partial \varepsilon'(\omega_{m})/\partial  \omega_{m}},
\end{equation}
where we restored the dielectric constant of surrounding medium $\varepsilon_{s}$. Note that the radiative decay rate for a spheroidal particle scales as the particle volume, implying higher radiation efficiency for larger particles. 
 
For  a spherical particle ($a=b$), we have $\varepsilon'(\omega_{sp})=-2$,  and so the plasmon radiative decay rate (\ref{mode-rate-rad-sp}) is recovered. The polarizability (\ref{polar-sp})   is recovered as well by using $U_{sp}= a^{3} \omega_{sp} [\partial \varepsilon'(\omega_{sp})/ \partial \omega_{sp}]/12$ and $ \bm{\mathcal{P}}_{sp}^{2}=a^{6}$, so that  
 \begin{equation}
 \label{factor}
\dfrac{\omega_{sp}\bm{\mathcal{P}}_{sp}^{2}}{4U_{sp}}=\dfrac{3a^{3}}{\partial \varepsilon'(\omega_{sp})/ \partial \omega_{sp}}.
 \end{equation}
For a nanosphere in a dielectric medium, the right-hand side of Eq.~(\ref{factor}) should be multiplied by $\varepsilon_{s}$.

\section{Mode volume and Purcell factor for spheroidal particles}

Using Gauss's law and expressing local fields in terms of potentials,  the  mode density projected along the nanorod major axis takes the form
\begin{equation}
\label{mode-density-proj1}
{\rho}_{L}(\bm{r})= 
\frac{2}{\omega_{m}\partial \varepsilon'(\omega_{m})/\partial  \omega_{m}}
\frac{\left [\nabla_{n}\Phi_{m}(\bm{r} )\right ]^{2}}{\int \!dS  \Phi_{m}\nabla_{n}\Phi_{m}},
\end{equation}
where integration takes place over the metal surface. For $\bm{r}$ at the distance $d$ from the tip of a prolate spheroidal particle  with major and minor semiaxes $a$ and $b$, respectively,  so that $\xi_{1}=a/\sqrt{a^{2}-b^{2}}$ at the surface, and using that $h_{\xi}=f$  along the $z$-axis, we obtain
\begin{equation}
\label{mode-density-proj2}
{\rho}_{L}= \dfrac{1}{\mathcal{V}_{L}}=\dfrac{2}{V_{\rm met}\,\omega_{L}}\left [\dfrac{\partial \varepsilon'(\omega_{L})}{\partial  \omega_{L}}\right ]^{-1}
%\frac{2}{\omega_{L}\partial \varepsilon'(\omega_{L})/\partial  \omega_{L}}
\left [\frac{Q'_{1}(\xi)\xi_{1}}{Q_{1}(\xi_{1})}\right ]^{2},
\end{equation}
where $\xi=(a+d)/\sqrt{a^{2}-b^{2}}$ and $V_{\rm met}=4\pi ab^{2}/3$ is the Au nanorod volume.  The plasmon frequency $\omega_{L}$ follows from the boundary condition $\varepsilon'(\omega_{L})=\varepsilon_{s} Q'_{1}(\xi_{1})\xi_{1}/Q_{1}(\xi_{1})$. In the limit of a spherical particle of radius $a$, i.e., $f\rightarrow 0$ and $\xi \rightarrow \infty$ as $b\rightarrow a$, we have $Q(\xi)\approx 1/3\xi^{2}$, yielding 
\begin{equation}
\label{mode-density-sphere-a}
{\rho}_{\rm sph}=\dfrac{1}{\mathcal{V}_{\rm sph}}
=
\dfrac{6}{\pi \omega_{L}}\left [\dfrac{\partial \varepsilon'(\omega_{L})}{\partial  \omega_{L}}\right ]^{-1}
 \frac{a^{3}}{(a+d)^{6}}.
\end{equation}
Note that for random dipole orientations, the orientational averaging results in the additional factor 1/3 in Eqs.~(\ref{mode-density-proj2}) and (\ref{mode-density-sphere-a}). Finally, the Purcell factor for a QE at distance $d$ from the nanorod tip   is given by 
\begin{equation}
\label{purcell-spheroid}
F_{p}= \dfrac{12\pi \varepsilon_{s}Q_{L}}{k^{3}V_{\rm met}\omega_{L}\partial \varepsilon'(\omega_{L})/\partial  \omega_{L} }
%\left [\dfrac{\partial \varepsilon'(\omega_{L})}{\partial  \omega_{L}}\right ]^{-1}
\left [\frac{Q'_{1}(\xi)\xi_{1}}{Q_{1}(\xi_{1})}\right ]^{2},
\end{equation}
and scales as $(k^{3}V_{\rm met})^{-1}$.

%%%%%%%%%%%%%%%%%%%%%%%%%%%%%%%%%%%%%


\begin{thebibliography}{99}

\bibitem{atwater-jap05} S. A. Maier and H. A. Atwater, 
%\textit{Plasmonics: Localization and guiding of electromagnetic energy in metal/dielectric structures},
J. Appl. Phys. \textbf{98}, 011101 (2005).

\bibitem{ozbay-science06} E. Ozbay, 
%\textit{Plasmonics: merging photonics and electronics at nanoscale dimensions},
Science  \textbf{311}, 189 (2006).

\bibitem{stockman-review} M. I. Stockman, 
%\textit{Nanoplasmonics: From Present into Future}, 
in \textit{Plasmonics: Theory and Applications}, edited by T. V. Shahbazyan and M. I. Stockman (Springer, New York, 2013).

%\bibitem{stockman-review} See, e.g., M. I. Stockman,
%%Nanoplasmonics: past, present, and glimpse into future
%Opt. Express \textbf{19},  22029  (2011).

\bibitem{sers} E. C. Le Ru and P. G. Etchegoin, \textit{Principles of Surface-Enhanced Raman Spectroscopy} (Elsevier, Oxford, 2009).

\bibitem{feldmann-prl02}E. Dulkeith, A. C. Morteani, T. Niedereichholz, T. A. Klar, J. Feldmann, S. A. Levi, F. C. J. M.. van Veggel, D. N. Reinhoudt, M. Moller, and D. I. Gittins, 
%Fluorescence quenching of dye molecules near gold nanoparticles: Radiative and nonradiative effects
Phys. Rev. Lett. \textbf{89}, 203002 (2002).

\bibitem{artemyev-nl02} O. Kulakovich, N. Strekal, A. Yaroshevich, S. Maskevich, S. Gaponenko, I. Nabiev, U. Woggon, and M. Artemyev,
%\textit{Enhanced Luminescence of CdSe Quantum Dots on Gold Colloids},
Nano Lett.  \textbf{2}, 1449 (2002).



\bibitem{novotny-prl06}P. Anger, P. Bharadwaj, and L. Novotny, 
%Enhancement and quenching of single-molecule fluorescence.
Phys. Rev. Lett. \textbf{96}, 113002 (2006).

\bibitem{sandoghdar-prl06}S. K\"{u}hn, U. Hakanson, L. Rogobete, and V. Sandoghdar, 
%Enhancement of Single-Molecule Fluorescence Using a Gold Nanoparticle as an Optical Nanoantenna
Phys. Rev. Lett. \textbf{97}, 017402 (2006).

\bibitem{halas-nl07} F. Tam, G. P. Goodrich, B. R. Johnson, and N. J. Halas, 
%Plasmonic Enhancement of Molecular Fluorescence
Nano Lett. \textbf{7}, 496 (2007).

\bibitem{halas-acsnano09}
R. Bardhan, N. K. Grady, J. R. Cole, A. Joshi, and N. J. Halas,
%\textit{Fluorescence Enhancement by Au Nanostructures: Nanoshells and Nanorods},
ACS Nano \textbf{3}, 744 (2009).

\bibitem{ming-nl09}
T. Ming, L. Zhao, Z. Yang, H. Chen, L. Sun, J. Wang, and C. Yan,
%\textit{Strong Polarization Dependence of Plasmon-Enhanced Fluorescence on Single Gold Nanorods},
Nano Lett. \textbf{9}, 3896 (2009).



%%%%%%%%%%%Strong Coupling


\bibitem{bellessa-prl04} J. Bellessa, C. Bonnand, J. C. Plenet, and J. Mugnier, 
%\textit{Strong Coupling between Surface Plasmons and Excitons in an Organic Semiconductor},
Phys. Rev. Lett. \textbf{93}, 036404 (2004).

\bibitem{sugawara-prl06} Y. Sugawara, T. A. Kelf, J. J. Baumberg, M. E. Abdelsalam, and P. N. Bartlett, 
%\textit{Strong Coupling between Localized Plasmons and Organic Excitons in Metal Nanovoids},
Phys. Rev. Lett. \textbf{97}, 266808 (2006).

\bibitem{wurtz-nl07} G. A. Wurtz, P. R. Evans, W. Hendren, R. Atkinson, W. Dickson, R. J. Pollard, A. V. Zayats, W. Harrison, and C. Bower,
%\textit{Molecular Plasmonics with Tunable Exciton−Plasmon Coupling Strength in J-Aggregate Hybridized Au Nanorod Assemblies},
 Nano Lett. \textbf{7}, 1297 (2007).

\bibitem{fofang-nl08} N. T. Fofang, T.-H. Park, O. Neumann, N. A. Mirin, P. Nordlander, and N. J. Halas, 
%\textit{Plexcitonic Nanoparticles: Plasmon−Exciton Coupling in Nanoshell−J-Aggregate Complexes},
Nano Lett. \textbf{8}, 3481 (2008).

\bibitem{hakala-prl09} T. K. Hakala, J. J. Toppari, A. Kuzyk, M. Pettersson, H. Tikkanen, H. Kunttu, and P. Torma, 
%\textit{Vacuum Rabi Splitting and Strong-Coupling Dynamics for Surface-Plasmon Polaritons and Rhodamine 6G Molecules},
Phys. Rev. Lett. \textbf{103}, 053602 (2009).

\bibitem{gomez-nl10} D. E. Gomez, K. C. Vernon, P. Mulvaney, and T. J. Davis, 
%\textit{Surface Plasmon Mediated Strong Exciton−Photon Coupling in Semiconductor Nanocrystals},
Nano Lett. \textbf{10}, 274 (2010).


\bibitem{manjavacas-nl11} A. Manjavacas, F. J. Garcia de Abajo, and P. Nordlander, 
%\textit{Quantum Plexcitonics: Strongly Interacting Plasmons and Excitons},
Nano Lett. \textbf{11}, 2318 (2011).

\bibitem{berrier-acsnano11} A. Berrier, R. Cools, C. Arnold, P. Oﬀermans, M. Crego-Calama, S. H. Brongersma, and J. Gomez-Rivas, 
%\textit{Active Control of the Strong Coupling Regime between Porphyrin Excitons and Surface Plasmon Polaritons,}
ACS Nano \textbf{5}, 6226 (2011).

\bibitem{salomon-prl12} A. Salomon, R. J. Gordon, Y. Prior, T. Seideman, and M. Sukharev, 
%\textit{Strong Coupling between Molecular Excited States and Surface Plasmon Modes of a Slit Array in a Thin Metal Film},
Phys. Rev. Lett. \textbf{109}, 073002 (2012).

\bibitem{guebrou-prl12} S. Aberra Guebrou, C. Symonds, E. Homeyer, J. C. Plenet, Y. N. Gartstein, V. M. Agranovich, and J. Bellessa, 
%\textit{Coherent Emission from a Disordered Organic Semiconductor Induced by Strong Coupling with Surface Plasmons},
Phys. Rev. Lett. 108, 066401 (2012).

%\bibitem{garcia-prl13}A. Gonzalez-Tudela, P. A. Huidobro, L. Martin-Moreno, C. Tejedor, and F. J. Garcia-Vidal, 
%\textit{Theory of Strong Coupling between Quantum Emitters and Propagating Surface Plasmons}, 
%Phys. Rev. Lett. \textbf{110}, 126801 (2013).

%\bibitem{gomez-jpcb13} D. E. Gomez, S. S. Lo, T. J. Davis, and G. V. Hartland, 
%\textit{Picosecond Kinetics of Strongly Coupled Excitons and Surface Plasmon Polaritons},
%J. Phys. Chem. B \textbf{117}, 4340 (2013).

\bibitem{antosiewicz-acsphotonics14} T. Antosiewicz, S. P. Apell, and T. Shegai, 
%\textit{Plasmon-Exciton Interactions in a Core-Shell Geometry: From Enhanced Absorption to Strong Coupling}, 
ACS Photonics, \textbf{1}, 454 (2014).

\bibitem{luca-apl14}A. De Luca, R. Dhama, A. R. Rashed, C. Coutant, S. Ravaine, P. Barois, M. Infusino, and G. Strangi,
%\textit{Double strong exciton-plasmon coupling in gold nanoshells infiltrated with fluorophores}, 
Appl. Phys. Lett. \textbf{104}, 103103 (2014).




%%%%%%%%%%%%%%%%%%%Spaser

\bibitem{bergman-prl03}
D. J. Bergman  and M. I. Stockman,
%\textit{Surface plasmon amplification by stimulated emission of radiation: quantum generation of coherent surface plasmons in nanosystems},
 Phys. Rev. Lett.,  \textbf{90}, 027402, (2003).
%%---
%
\bibitem{stockman-natphot08}
M. I. Stockman,
%\textit{Spasers explained},
Nat. Photonics \textbf{2}, 327, (2008).
%%---

%
\bibitem{noginov-nature09}
M. A. Noginov,  G. Zhu, A. M. Belgrave,  R. Bakker, V. M. Shalaev, E. E. Narimanov, S. Stout,  E. Herz, T. Suteewong  and  U. Wiesner,  
%\textit{Demonstration of a spaser-based nanolaser},
Nature, \textbf{460}, 1110, (2009).


%
%%%%%%%Resonators
%
%
\bibitem{carminati-oc06} R. Carminati, J. J. Greffet, C. Henkel, J. M. Vigoureux, 
%Radiative and non-radiative decay of a single molecule close to a metallic nanoparticle
Opt. Commun. \textbf{261} 368 (2006).


\bibitem{greffet-prl10} J.-J. Greffet, M. Laroche, and F. Marquier,
%Impedance of a Nanoantenna and a Single Quantum Emitter,
Phys. Rev. Lett. \textbf{105}, 117701 (2010).

\bibitem{lalanne-prl13} C. Sauvan, J. P. Hugonin, I. S. Maksymov, and P. Lalanne,
%Theory of the Spontaneous Optical Emission of Nanosize Photonic and Plasmon Resonators,
Phys. Rev. Lett. \textbf{110}, 237401 (2013).

\bibitem{hughes-njp14}R.-C. Ge, P. T. Kristensen, J. F. Young, and S. Hughes, 
%Quasinormal mode approach to modelling lightemission and propagation in nanoplasmonics
New J. Phys. \textbf{16}, 113048 (2014).

\bibitem{pelton-np15} M. Pelton,
%Modified spontaneous emission in nanophotonic structures
Nat. Photonics \textbf{9}, 427 (2015).


\bibitem{belov-sr15} A. E. Krasnok, A. P. Slobozhanyuk, C. R. Simovski, S. A. Tretyakov, A. N. Poddubny, A. E. Miroshnichenko, Y. S. Kivshar, and  P. A. Belov,
%Antenna model of the Purcell effect
Sci. Rep. \textbf{5},  12956 (2015).

\bibitem{derex-jo16} G. C. des Francs, J Barthes, A Bouhelier, J C Weeber, A Dereux, A Cuche, and C Girard, 
%Plasmonic Purcell factor and coupling efficiency to surface plasmons. Implications for addressing and controlling optical nanosources,
J. Opt. \textbf{18} 094005 (2016).

\bibitem{greffet-acsph17} F. Marquier, C. Sauvan, and J.-J. Greffet,  
%\textit{Revisiting Quantum Optics with Surface Plasmons and Plasmonic Resonators},
ACS Phot. \textbf{4}, 2091–2101 (2017).

\bibitem{koenderink-acsphot17}  A. F. Koenderink, 
%\textit{Single-Photon Nanoantennas}, 
ACS Photonics \textbf{4}, 710 (2017). DOI: 10.1021/acsphotonics.7b00061


\bibitem{lalanne-prb18}  W. Yan, R.i Faggiani, and P. Lalanne,
%Rigorous modal analysis of plasmonic nanoresonators
Phys. Rev. B \textbf{97}, 205422 (2018).

%%%%%%%%%%%%%%%%%%%%%%
\bibitem{novotny-book} L. Novotny and B. Hecht, \textit{Principles of Nano-Optics} (CUP, New York, 2012).

\bibitem{purcell-pr46} E. M. Purcell, 
%Spontaneous emission probabilities at radio frequencies
Phys. Rev. \textbf{69},  681 (1946).

\bibitem{maier-oe06} S. Maier, 
%Plasmonic ﬁeld enhancement and SERS in the effective mode volume picture
Opt. Express 14, 1957 (2006).

\bibitem{koenderink-ol10} A. F. Koenderink, 
%On the use of Purcell factors for plasmon antennas
Opt. Lett. \textbf{35} 4208 (2010).

\bibitem{stefano-jpcm10} O. Di Stefano, N. Fina, S. Savasta, R. Girlanda, and M. Pieruccini,
%Calculation of the local optical density of states in absorbing and gain media
J. Phys.: Condens. Matter \textbf{22}, 315302 (2010).

\bibitem{andreani-prb12} F. Alpeggiani, S. D'Agostino, and L. C. Andreani,
%Surface plasmons and strong light-matter coupling in metallic nanoshells
Phys. Rev. B \textbf{86}, 035421 (2012).


\bibitem{hughes-ol12} P. T. Kristensen, C. Van Vlack, and S. Hughes,
%Generalized effective mode volume for leaky optical cavities
Opt. Lett. \textbf{37},  1649  (2012).

\bibitem{hughes-acsphot14} P. T. Kristensen  and S. Hughes,
%Modes and Mode Volumes of Leaky Optical Cavities and Plasmonic Nanoresonators
ACS Phot. \textbf{1}, 2 (2014).

\bibitem{muljarov-prb16}E. A. Muljarov and W. Langbein,
%Exact mode volume and Purcell factor of open optical systems
Phys. Rev. B \textbf{94}, 235438 (2016).


\bibitem{lalanne-pra14} C. Sauvan, J. P. Hugonin, R. Carminati, and P. Lalanne,
%Modal representation of spatial coherence in dissipative and resonant photonic systems
Phys. Rev. A \textbf{89}, 043825 (2014).

\bibitem{hughes-pra15} P. T. Kristensen, R.-C. Ge, and S. Hughes,
%Normalization of quasinormal modes in leaky optical cavities and plasmonic resonators
Phys. Rev. A \textbf{92}, 053810 (2015).


\bibitem{bergman-17}P. Y. Chen, D. J. Bergman, Y. Sivan,
%Generalizing normal mode expansion of electromagnetic Green's tensor to lossy resonators in open systems
arXiv:1711.00335.

\bibitem{lalanne-lpr18}  P. Lalanne,  W. Yan,  K. Vynck,  C. Sauvan, and  J.‐P. Hugonin,
%Light Interaction with Photonic and Plasmonic Resonances
Laser Photon. Rev.  \textbf{12}, 1700113 (2018).





%Controversy










%\bibitem{manjavacas-nl11} A. Manjavacas, F. J. Garcia de Abajo, and P. Nordlander, 
%%\textit{Quantum Plexcitonics: Strongly Interacting Plasmons and Excitons},
%Nano Lett. \textbf{11}, 2318 (2011).
%
%\bibitem{garcia-prl14} A. Delga, J. Feist, J. Bravo-Abad, and F. J. Garcia-Vidal,
%%\textit{Quantum emitters near a metal nanoparticle: strong coupling and quenching},
%Phys. Rev. Lett. \textbf{112}, 253601 (2014).

%\bibitem{stockman-jo10} M. I. Stockman,
%%\textit{Spaser as Nanoscale Quantum Generator and Ultrafast Amplifier},
%J. Opt. \textbf{12}, 024004, (2010).

%%%%%%%%%%%%%%%%%%%%%%%%%%%%





%%%%%%%%%%%%%%%%%%%%%%%%%%%%%%%%%%%%


%\bibitem{mahan-book} See, e.g., G. D. Mahan, {\it Many-Particle Physics} (Plenum, New York, 1990).






\bibitem{shahbazyan-prl16} T. V. Shahbazyan, 
%Local density of states for nanoplasmonics
Phys. Rev. Lett. \textbf{117}, 207401 (2016).

\bibitem{landau} L. D. Landau and E. M. Lifshitz, {\it Electrodynamics of Continuous Media}  (Elsevier, Amsterdam, 2004). 


\bibitem{optical} W. Vogel and  D.‐G. Welsch, \textit{Quantum Optics} (Wiley, 2016).


\bibitem{poddubny-prl16} A. N. Poddubny, I. V. Iorsh, and A. V. Sukhorukov, 
%Generation of Photon-Plasmon Quantum States in Nonlinear Hyperbolic Metamaterials
Phys. Rev. Lett. \textbf{117}, 123901 (2016) and Supplemental Material.


\bibitem{shahbazyan-prb16} T. V. Shahbazyan, 
%Landau damping of surface plasmons in metal nanostructures
Phys. Rev. B \textbf{94}, 235431 (2016).

\bibitem{mortensen-nc14} N. A. Mortensen,	S. Raza,	M. Wubs,	T. S{\o}ndergaard, and S. I. Bozhevolnyi, 
%\textit{A generalized non-local optical response theory for plasmonic nanostructures},
Nat. Commun. \textbf{5}, 3809 (2014).

\bibitem{mortensen-optica17} M. K. Dezfouli, C. Tserkezis, N. A. Mortensen, and S. Hughes, Optica \textbf{4}, 1503 (2017).




\end{thebibliography}
\end{document}